\documentclass[reviewcopy]{elsart}

\usepackage{graphicx}

\usepackage{amssymb,epsf,latexsym,hyperref}

\newcommand{\dH}{de~Haas-van~Alphen }
\newcommand{\mz}{\mathfrak Z}
\newcommand{\csch}{\mbox{\rm csch}}

\begin{document}

\begin{frontmatter}

\title{Ideal Fermi gases in harmonic oscillator potential traps}

\author{David J. Toms}
\ead{d.j.toms@newcastle.ac.uk}
\ead[url]{http://www.staff.ncl.ac.uk/d.j.toms/}

\address{School of Mathematics and Statistics, University of Newcastle Upon Tyne,\\
Newcastle Upon Tyne, United Kingdom NE1 7RU}

\begin{abstract}
We study the thermodynamic properties of an ideal gas of fermions
in a harmonic oscillator confining potential. The analogy between
this problem and the de~Haas-van~Alphen effect is discussed and
used to obtain analytical results for the chemical potential and
specific heat in the case of both isotropic and anisotropic
potentials. Step-like behaviour in the chemical potential, first
noted in numerical studies, is obtained analytically and shown to
result in an oscillatory behaviour of the specific heat when the
particle number is varied. The origin of these oscillations is
that part of the thermodynamic potential responsible for the
de~Haas-van~Alphen-type effect. At low temperatures we show analytically that there are significant deviations in the specific heat from the
expected linear temperature dependence, again as a consequence of
the de~Haas-van~Alphen part of the thermodynamic potential.
Results are given for one, two, and three spatial dimensions. In the anisotropic case we show how the specific heat jumps as the ratio of oscillator frequencies varies.
\end{abstract}

\end{frontmatter}

\section{\label{sec1}Introduction}

The experimental realization~\cite{BEC} of Bose-Einstein
condensation in confined gases of atoms at low temperatures has
been the stimulus for a wide range of experimental and theoretical
investigations. (See Ref.~\cite{PethickSmith} for a review.) By
ingenious experimental techniques it is possible to prepare the
atoms as either bosons or fermions. There have now been a number
of experiments on trapped gases of fermions, as well as mixtures
of bosons and fermions~\cite{fermi}.

For a gas of fermions, it is a good approximation to model the
system as a collection of non-interacting particles obeying
Fermi-Dirac statistics confined by a simple harmonic oscillator
potential. Unlike the case of bosons, the dominant $s$-wave
scattering channel is suppressed making the effects of
interactions less important~\cite{BR}. With this simple
theoretical model for the system, it is easy to obtain exact and
simple expressions for the single-particle energy levels using the
standard quantum mechanical result for the simple harmonic
oscillator. This can serve as a starting point for a theoretical
analysis of the thermodynamic behaviour of the system.

Because the energy levels for a set of particles (bosons or
fermions) confined by a simple harmonic oscillator potential form
a discrete set, in order to calculate the thermodynamic properties
we must be able to perform sums over the quantum numbers labelling
the states; this can lead to difficulties in evaluation. By
restricting the parameters of the system (temperature and
oscillator frequencies for example) it may be possible to argue
that the sums can be approximated with integrals, an approximation
that renders the computation of analytical results easier. We will
refer to the approximation of sums with integrals as the continuum
approximation, since it is equivalent to regarding the discrete
energy spectrum as continuous. This approximation was first
applied to the Fermi gas in the now classic paper of Butts and
Rokhsar~\cite{BR}. Even when one goes beyond this approximation,
as we will show below, the continuum approximation can still be
very accurate in some temperature regimes, and can capture the
leading order behaviour if the temperature is not too low.

The first study that showed there might be more features present
beyond what is obtained by the continuum approximation was given
by Schneider and Wallis~\cite{SW}. These authors computed the
thermodynamic expressions numerically, and their results showed
that there were a number of step-like features present, not
predicted by the continuum approximation. In particular, the
chemical potential, which is a smooth function in the continuum
approximation, has a series of step-like jumps when a numerical
analysis of the exact result is performed. This in turn can lead
to step-like, or oscillatory, behaviour of other thermodynamic
properties, such as the specific heat. Another important
consequence of the work presented in Ref.~\cite{SW} is that at
sufficiently low temperatures there are significant deviations in
the specific heat from the linear temperature dependence
predicted~\cite{BR} by the continuum approximation. The main
purpose of the present paper is to analyze the situation
analytically to obtain the step-like behaviour of the chemical
potential, and to apply this method to the evaluation of the
specific heat.

In addition to the two studies already mentioned~\cite{BR,SW},
there have been a number of other related theoretical works.
Ref.~\cite{BB} generalized the calculations of Ref.~\cite{SW} to a
gas with two spin states with an interaction between the two
states. (We will make some brief comments on the effects of
interactions in Sec.~\ref{sec5}.) A number of authors
\cite{Brosens,VMT,Gleisberg,BvZ,Wang,vZetal,BM,vZ} have obtained
analytical approximations in a number of cases using path
integral, Green's function, or density matrix methods. The approach described in the present paper was used to study partially confined gases in Ref.~\cite{DJTpartial}

In the present paper we will concentrate on the behaviour of the
chemical potential and specific heat, and will use the evaluation
of the thermodynamic potential as the basic starting point. The
key observation is that for a simple harmonic oscillator confining
potential, the evaluation of the thermodynamic potential is
mathematically very similar to that of a gas of charged fermions
in a constant magnetic field. Landau~\cite{Landau} was the first
to show that the thermodynamic properties of electrons in a
constant magnetic field undergo oscillations whose period is
determined by the inverse of the magnetic field strength. This
gave a theoretical basis for the observations of de Haas and van
Alphen, now referred to as the \dH effect. (See
Ref.~\cite{Ashcroft} for a good discussion of the
de~Haas-van~Alphen effect and the important role it plays in
condensed matter physics.) What we will show here is that for the
trapped Fermi gas the step-like behaviour found in Ref.~\cite{SW}
is completely analogous to the \dH effect. Just as in the \dH
effect, where use of the continuum approximation would miss the
oscillations, here too we must go beyond the continuum
approximation.

The outline of our paper is as follows. In Sec.~\ref{sec2} we
describe the general method following the classic analysis of
Sondheimer and Wilson~\cite{Sond}. There are some differences with
the \dH case considered in Ref.~\cite{Sond} that we will describe.
We will show how the continuum approximation comes about in the
method, and even find the next order correction to it, beyond the
leading order result given in Ref.~\cite{BR}. In Sec.~\ref{sec3}
we analyze the \dH contribution to the thermodynamic potential for
a 3-dimensional isotropic harmonic oscillator potential, and
obtain approximate analytical results for the chemical potential.
This is then used to examine the specific heat. Results for the 1-
and 2-dimensional gases are given in Sec.~\ref{sec4} and the
results compared with the 3-dimensional case. In Sec.~\ref{sec2anis} we analyze the situation when the trapping potential is anisotropic. Finally,
Sec.~\ref{sec5} contains a brief summary and a short discussion of
the main results.

\section{\label{sec2}General method}
\setcounter{equation}{0}

We will begin with the grand canonical ensemble and the
thermodynamic potential $\Omega$ for a system of fermions with energy levels $E_n$ defined by
\begin{equation}
\Omega=-T\sum_nf(E_n)\;,\label{2.1}
\end{equation}
where
\begin{equation}
f(E)=\ln\left\lbrack1+\exp(-\beta(E-\mu))\right\rbrack\;.\label{2.2}
\end{equation}
We use the usual notation $\beta=1/T$, with $T$ the temperature,
and choose units such that the Boltzmann constant is equal to one.
Take the Laplace transform of $f(E)$, and call it
$\varphi(\beta)$, so that
\begin{equation}
\varphi(\beta)=\int\limits_{0}^{\infty} dE\,e^{-\beta
E}f(E)\;.\label{2.3}
\end{equation}
The inverse Laplace transform of Eq.~(\ref{2.3}) reads
\begin{equation}
f(E)=\frac{1}{2\pi
i}\int\limits_{c-i\infty}^{c+i\infty}d\beta\,e^{\beta
E}\varphi(\beta)\;,\label{2.4}
\end{equation}
where $c$ is an arbitrary constant chosen so that the integration
path lies to the right of any singularities of $\varphi(\beta)$.
We can now use Eq.~(\ref{2.4}) in Eq.~(\ref{2.1}). The sum over
the energy levels that results can be related to the partition
function for the $\mu=0$ system, defined by
\begin{equation}
Z(\beta)=\sum_ne^{-\beta E_n}\;.\label{2.5}
\end{equation}
We can then write $\Omega$ defined by Eq.~(\ref{2.1}) as
\begin{equation}
\Omega=-\frac{T}{2\pi i}
\int\limits_{c-i\infty}^{c+i\infty}d\beta\,Z(-\beta)\varphi(\beta)\;.\label{2.6}
\end{equation}
It is necessary here to regard $Z(\beta)$ as a function of $\beta$
with $\beta$ viewed as a complex variable. $Z(-\beta)$ is the
result obtained from this complex function by analytic
continuation of the definition Eq.~(\ref{2.5}) to $\Re(\beta)<0$.

We now define $\mz(E)$ to be the Laplace transform of
$\beta^{-2}Z(\beta)$. (The factor of $\beta^{-2}$ is for later
convenience~\cite{Sond}.) This results in the definition
\begin{equation}
Z(\beta)=\beta^2\int\limits_{0}^{\infty}dE\,e^{-\beta E}\mz(E)\;.
\label{2.7}
\end{equation}
The inverse Laplace transform gives
\begin{equation}
\mz(E)=\frac{1}{2\pi
i}\int\limits_{c-i\infty}^{c+i\infty}d\beta\,e^{\beta
E}\beta^{-2}Z(\beta) \;.\label{2.8}
\end{equation}
Substitution of Eq.~(\ref{2.7}) into Eq.~(\ref{2.6}) followed by
the use of Eq.~(\ref{2.4}) results in
\begin{equation}
\Omega=-T\int\limits_{0}^{\infty}dE\,\mz(E)\frac{\partial^2f(E)}{\partial
E^2} \;.\label{2.9}
\end{equation}
The Fermi-Dirac distribution function $F(E)$ is
\begin{eqnarray}
F(E)&=&\left\lbrack e^{\beta(E-\mu)}+1 \right\rbrack^{-1}\label{2.10}\\
&=&-T\frac{\partial}{\partial E}f(E) \;,\label{2.11}
\end{eqnarray}
if we use Eq.~(\ref{2.2}). This leads to
\begin{equation}
\Omega=\int\limits_{0}^{\infty}dE\,\mz(E) \frac{\partial}{\partial
E}F(E) \;,\label{2.12}
\end{equation}
giving the key starting point in the Sondheimer-Wilson~\cite{Sond}
analysis of the \dH effect. The method rests on an evaluation of
$\mz(E)$ defined by Eq.~(\ref{2.7}) or Eq.~(\ref{2.8}) and its use
in Eq.~(\ref{2.12}).

To evaluate $\mz(E)$ in Eq.~(\ref{2.8}) we require the partition
function, or at least some information about the singularities of
$\beta^{-2}Z(\beta)$ in the complex $\beta$-plane. For the case of
a simple harmonic oscillator potential this information is very
easy to obtain. We will consider the $D$-dimensional oscillator
potential
\begin{equation}
V(\mathbf{x})=\frac{1}{2}m\sum_{j=1}^{D}\omega_j^2x_j^2
\;,\label{2.13}
\end{equation}
with $\mathbf{x}=(x_1,\ldots,x_D)$ the $D$ spatial coordinates.
The energy levels are simply given by (in $\hbar=1$ units)
\begin{equation}
E_{\mathbf{n}}=\sum_{j=1}^{D}\left( n_j+\frac{1}{2}
\right)\omega_j \;,\label{2.14}
\end{equation}
where $\mathbf{n}=(n_1,\ldots,n_D)$ and $n_j=0,1,2,\ldots$ for
$j=1,\ldots,D$. We then have $Z(\beta)$ in Eq.~(\ref{2.5})
expressed as a product of geometric series that are easily summed
to give
\begin{equation}
Z(\beta)=\prod_{j=1}^{D}\frac{e^{-\beta\omega_j/2}}{(1-e^{-\beta\omega_j}
)} \;.\label{2.15}
\end{equation}
Because of the neglect of interactions, and the simple form of the
potential, it has been possible to evaluate the partition function
in closed form.

To obtain $\mz(E)$ we need to know the singularity structure of
$\beta^{-2}Z(\beta)$. It is now obvious from Eq.~(\ref{2.15}) that
$\beta^{-2}Z(\beta)$ is a meromorphic function. There is a pole at
$\beta=0$ of order $D+2$ as well as a series of poles along the
positive and negative imaginary $\beta$-axis at
$\beta=\beta_{k_j}=2\pi ik_j/\omega_j$ with
$k_j=\pm1,\pm2,\ldots$. The order of these poles depends on the
relative ratios of the harmonic oscillator frequencies. If none of
the oscillator frequencies are rational multiples of any of the
others, all of the poles, apart from the one at $\beta=0$, will be
simple. However, if some of the oscillator frequencies are
rational multiples of the others, some of the poles away from the
origin will be higher order. This makes a determination of the
residues of poles along the positive and negative imaginary
$\beta$-axis for a general harmonic oscillator potential an
inelegant treatment of special cases. We will return to this
problem later.

We can write $\mz(E)$ in Eq.~(\ref{2.8}) as
\begin{equation}
\mz(E)=\mz_0(E)+\mz_r(E) \;,\label{2.16}
\end{equation}
where we close the contour in the left hand side of the complex
plane and use $\mz_0(E)$ to denote the contribution to the
integral from the pole at $\beta=0$, and $\mz_r(E)$ the
contribution coming from the rest of the poles along the positive
and negative imaginary $\beta$-axis. When Eq.~(\ref{2.16}) is used
in Eq.~(\ref{2.12}) we obtain the thermodynamic potential as
\begin{equation}
\Omega=\Omega_0+\Omega_r \label{2.17}
\end{equation}
in an obvious way. As in the \dH effect, the oscillations in
thermodynamic quantities will come from $\Omega_r$, but we will
first examine $\Omega_0$. $\Omega_r$ will be evaluated in Sec.~\ref{sec3} for the isotropic potential, and in Sec.~\ref{sec2an} for the anisotopic potential.

To obtain $\mz_0(E)$ we need the residue of $\beta^{-2}e^{\beta
E}Z(\beta)$ at $\beta=0$. For general values of $D$ and arbitrary
frequencies $\omega_j$ this is difficult to write down in any
simple way. We will concentrate on just the cases $D=1,2,3$ here,
although there is no intrinsic reason why the method cannot be
extended to other values. A straightforward calculation yields
\begin{eqnarray}
\mz_0(E)&=&\frac{E^2}{2\omega}-\frac{\omega}{24}\ {\rm for}\
D=1,\label{2.18a}\\
\mz_0(E)&=&\frac{E}{24\omega_1\omega_2}  \left(
4E^2-\omega_1^2-\omega_2^2 \right)\ {\rm for}\
D=2,\label{2.18b}\\
\mz_0(E)&=&\frac{1}{5760\omega_1\omega_2\omega_3} \Big\lbrack
240E^4-120E^2(\omega_1^2+\omega_2^2+\omega_3^2)+ 7(\omega_1^4+\omega_2^4+\omega_3^4)\nonumber\\
&&+10(\omega_1^2\omega_2^2 +\omega_2^2\omega_3^2
+\omega_3^2\omega_1^2) \Big\rbrack \ {\rm for}\ D=3.\label{2.18c}
\end{eqnarray}
These results can now be used in Eq.~(\ref{2.12}) to obtain
\begin{equation}
\Omega_0=\int\limits_{0}^{\infty}dE\,\mz_0(E)
\frac{\partial}{\partial E}F(E) \;.\label{2.19}
\end{equation}

As $T\rightarrow0$, meaning that $\beta\mu\rightarrow\infty$, the
Fermi-Dirac distribution function approaches a step-function,
$F(E)\rightarrow\theta(\mu-E)$, so that $\frac{\partial}{\partial
E} F(E)\rightarrow-\delta(E-\mu)$. This crude approximation
results in
\begin{equation}
\Omega_0\rightarrow-\mz_0(\mu)\;,\label{2.20}
\end{equation}
and is really the first term in a systematic expansion due to
Sommerfeld. (See Refs.~\cite{Ashcroft,Huang,Pathria} for three
derivations of the Sommerfeld expansion.) In our case we find
\begin{equation}
\Omega_0\simeq-\mz_0(\mu)-\frac{\pi^2}{6}T^2\mz_0''(\mu)-
\frac{7\pi^4}{360}T^4\mz_0''''(\mu)\;. \label{2.21}
\end{equation}
Although the general Sommerfeld expansion contains higher order
terms, these vanish here because $\mz_0(E)$ is no more than
quartic in $E$ due to our restriction that $D\le3$. For arbitrary
$D$, the expansion Eq.~(\ref{2.21}) will contain more terms with
increasing powers of $T$. For low values of $T$ these should be
less important than those indicated in Eq.~(\ref{2.21}) in any
case. Using Eqs.~(\ref{2.18a}-\ref{2.18c}) in Eq.~(\ref{2.21})
results in
\begin{eqnarray}
\Omega_0&\simeq&-\frac{1}{2\omega}(\mu^2+\frac{\pi^2}{3}T^2)+\frac{\omega}{24}
\;,\
{\rm for\ }D=1\;,\label{2.22a}\\
\Omega_0&\simeq&-\frac{\mu^3}{6\omega_1\omega_2}+\frac{\mu}{24\omega_1\omega_2}
(\omega_1^2+\omega_2^2-4\pi^2 T^2)\;,\
{\rm for\ }D=2\;,\label{2.22b}\\
\Omega_0&\simeq&-\frac{\mu^4}{24\omega_1\omega_2\omega_3}
+\frac{\mu^2}{48\omega_1\omega_2\omega_3}(\omega_1^2+\omega_2^2+\omega_3^2-4\pi^2
T^2)\nonumber\\
&&-\frac{1}{5760\omega_1\omega_2\omega_3 }
\Big\lbrack7(\omega_1^4+\omega_2^4+\omega_3^4)
+10(\omega_1^2\omega_2^2+\omega_2^2\omega_3^2+\omega_3^2\omega_1^2
)\nonumber\\
&&-40\pi^2T^2
(\omega_1^2+\omega_2^2+\omega_3^2)+112\pi^4T^4\Big\rbrack\;,\ {\rm
for\ }D=3\;.\label{2.22c}
\end{eqnarray}

The thermodynamic properties of the system follow from a knowledge
of the thermodynamic potential. In particular, the average
particle number $N$ is given by
\begin{equation}
N=-\left.\left( \frac{\partial\Omega}{\partial\mu}
\right)\right|_{T,\omega} \;,\label{2.23}
\end{equation}
and the internal energy $U$ is given by
\begin{equation}
U=\left. \frac{\partial}{\partial\beta}(\beta\Omega)
\right|_{\beta\mu,\omega} \;.\label{2.24}
\end{equation}
In the most physically interesting case we take $D=3$. (The cases
of $D=1,2$ will be given later in Sec.~\ref{sec4}.) If we
temporarily ignore the contribution $\Omega_r$ to $\Omega$, we
have $N\simeq N_0$ where
\begin{eqnarray}
N_0&=&-\left.\left( \frac{\partial\Omega_0}{\partial\mu}
\right)\right|_{T,\omega}\label{2.25a}\\
&\simeq&\frac{\mu^3}{6\omega_1\omega_2\omega_3}-\frac{\mu}{24\omega_1\omega_2\omega_3}
(\omega_1^2+\omega_2^2+\omega_3^2-4\pi^2 T^2) \;.\label{2.25}
\end{eqnarray}
For large values of $\mu$, if we keep only the term in
Eq.~(\ref{2.25}) of order $\mu^3$, we find (with $N\simeq N_0$),
\begin{equation}
\mu^3\simeq 6\omega_1\omega_2\omega_3 N \;,\label{2.26}
\end{equation}
showing that $\mu\propto N^{1/3}$. For large values of $N$ we will
have $\mu$ much larger than the average oscillator frequency,
consistent with the assumptions made in deriving these results. It
is easy to obtain the next order correction to Eq.~(\ref{2.26}),
\begin{equation}
\frac{\mu^3}{\omega_1\omega_2\omega_3}\simeq6N\left\lbrace 1 +
\frac{(\omega_1^2+\omega_2^2+\omega_3^2-4\pi^2T^2)}{4(6N\omega_1\omega_2\omega_3)^{2/3}}
\right\rbrace \;,\label{2.27}
\end{equation}
showing that the correction to Eq.~(\ref{2.26}) becomes
increasingly unimportant for large values of $N$. If we use
Eq.~(\ref{2.27}) and rewrite the result in terms of the Fermi
energy defined by
\begin{equation}
E_F=\mu(T=0) \;,\label{2.28}
\end{equation}
it is easy to see that
\begin{equation}
\mu\simeq E_F\left(1- \frac{\pi^2 T^2}{3E_F^2} \right)
\;.\label{2.29}
\end{equation}
This agrees with the result in Ref.~\cite{BR} who used the
continuum approximation and only the $T$-dependent part of the
second term in Eq.~(\ref{2.25}). We have demonstrated here that
Eq.~(\ref{2.29}) holds true even when the next order approximation
for the density of states is included in the continuum
approximation. However, as we will see in the next section there
are important corrections to this from terms that come from beyond
the use of the continuum approximation and support the numerical
investigations of Ref.~\cite{SW}.

We can also calculate the contribution to the internal energy,
that we call $U_0$, coming from $\Omega_0$. Using
Eq.~(\ref{2.22c}) in Eq.~(\ref{2.24}) and then eliminating $\mu$
in favour of $N$ with Eq.~(\ref{2.27}) results in the following
approximation~:
\begin{equation}
U_0\simeq\frac{3}{4}(6\omega_1\omega_2\omega_3)^{1/3}N^{4/3}
+\frac{(\omega_1^2+\omega_2^2+\omega_3^2+4\pi^2T^2)}{8(6\omega_1\omega_2\omega_3)^{1/3}}\,N^{2/3}
\;,\label{2.30}
\end{equation}
assuming that $N$ is large. (The intermediate result for $U_0$ in
terms of $\mu$ can be found in Eq.~(\ref{3.24}) below for the
isotropic potential.)

The specific heat can be found using
\begin{equation}
C=\left.\left(\frac{\partial U}{\partial T}
\right)\right|_{N,\omega} \;.\label{2.31}
\end{equation}
The contribution coming from the approximate result in
Eq.~(\ref{2.30}) is then seen to be
\begin{equation}
C\simeq\pi^2(6\omega_1\omega_2\omega_3)^{-1/3}N^{2/3}T
\;.\label{2.32}
\end{equation}
As before this agrees with the result quoted in Ref.~\cite{BR} demonstrating a linear dependence on the temperature.
Again we will find that at low temperatures the part of $\Omega$
that has not yet been considered, namely $\Omega_r$, alters this
expected behaviour, confirming the numerical results of
Ref.~\cite{SW}. We consider the evaluation of $\Omega_r$ and its
effect in the next section.

\section{\label{sec3}Isotropic 3-dimensional potential}\setcounter{equation}{0}
\subsection{\label{sec3.1}Thermodynamic potential}

We will examine the case of the isotropic harmonic oscillator
potential with
\begin{equation}
\omega_1=\omega_2=\omega_3=\omega \;,\label{3.1}
\end{equation}
in which case the results of the previous section simplify. Our
aim here is to calculate $\Omega_r$ that has been neglected up to
now from the analysis. To do this we need to evaluate the
contribution to $\Omega$ coming from the poles of
$\beta^{-2}e^{\beta E}Z(\beta)$ along the imaginary $\beta$-axis
away from $\beta=0$. From Eq.~(\ref{2.15}), because
\begin{equation}
Z(\beta)=\frac{e^{-3\beta\omega/2}}{(1-e^{-\beta\omega})^{3}}
\;,\label{3.2a}
\end{equation}
there will be poles of order 3 at $\beta=\beta_k$ defined by
\begin{equation}
\beta_k=\frac{2\pi i}{\omega}\,k \;\label{3.2}
\end{equation}
for $k=\pm1,\pm2,\ldots$. It is a straightforward matter to
evaluate the residues of $\beta^{-2}e^{\beta E}Z(\beta)$ at
$\beta=\beta_k$ and show that their contribution to
Eq.~(\ref{2.8}) gives
\begin{eqnarray}
\mz_r(E)&=&\sum_{k=1}^{\infty}\frac{(-1)^k\omega}{16\pi^4k^4}
\left(6+\pi^2k^2-\frac{4\pi^2k^2 E^2}{\omega^2} \right) \cos(2\pi
k E/\omega)\nonumber\\
&&+\sum_{k=1}^{\infty}\frac{(-1)^k E}{2\pi^3k^3}
\sin(2\pi k E/\omega) \;.\label{3.3}
\end{eqnarray}
It now remains to use this in Eq.~(\ref{2.12}) and to try to
extract something useful from the result.

We have
\begin{equation}
\Omega_r=-\frac{\beta\omega}{64\pi^4}
\sum_{k=1}^{\infty}\frac{(-1)^k}{k^4}A_k- \frac{\beta}{8\pi^3}
\sum_{k=1}^{\infty}\frac{(-1)^k}{k^3}B_k \;,\label{3.4}
\end{equation}
where
\begin{eqnarray}
A_k&=&\int\limits_{0}^{\infty}dE \left(6+\pi^2k^2-\frac{4\pi^2k^2
E^2}{\omega^2} \right)\frac{\cos(2\pi k E/\omega)}{\cosh^2\lbrack
\frac{1}{2}\beta(E-\mu) \rbrack} \;,\label{3.5}\\
B_k&=&\int\limits_{0}^{\infty}dE\;E\; \frac{\sin(2\pi k
E/\omega)}{\cosh^2\lbrack \frac{1}{2}\beta(E-\mu) \rbrack}
\;.\label{3.6}
\end{eqnarray}
In these last two integrals we can make the change of variable
\begin{equation}
E=\mu+\frac{2}{\beta}\,\theta \;,\label{3.7}
\end{equation}
defining a new integration variable $\theta$. The lower limits on the
integrals in Eqs.~(\ref{3.5}) and (\ref{3.6}) then become
$-\beta\mu/2$. If we look at low enough temperatures, specifically
$T<<\mu$ as we have already assumed, then to a good approximation
(up to exponentially small terms) we can replace the lower limits
on the integrals defining $A_k$ and $B_k$ with $-\infty$. The
approximate results then become
\begin{eqnarray}
A_k&\simeq&\frac{2}{\beta}\int\limits_{-\infty}^{\infty}\frac{d\theta}{\cosh^2\theta}
\left\lbrack 6+\pi^2k^2-\frac{4\pi^2k^2\mu^2}{\omega^2}
-\frac{16\pi^2k^2\mu\theta}{\beta\omega^2}
-\frac{16\pi^2k^2\theta^2}{\beta^2\omega^2} \right\rbrack
\nonumber\\
&&\qquad\qquad\times \cos\left(\frac{2\pi k\mu}{\omega}+\frac{4\pi
k\theta}{\beta\omega} \right) \;,\label{3.8}\\
B_k&\simeq&\frac{2}{\beta}\int\limits_{-\infty}^{\infty}\frac{d\theta}{\cosh^2\theta}
\left( \mu+\frac{2\theta}{\beta}\right) \sin\left(\frac{2\pi
k\mu}{\omega}+\frac{4\pi k\theta}{\beta\omega} \right)
\;.\label{3.9}
\end{eqnarray}
The integrals in Eqs.~(\ref{3.8}) and (\ref{3.9}) may be evaluated
exactly using residues. All of the expressions required may be
related to the basic integral
\begin{equation}
\int\limits_{-\infty}^{\infty}\frac{\cos(a\theta+b)}{\cosh^2\theta}
 \;d\theta =\frac{\pi a\cos b}{\sinh(\frac{\pi}{2}a)} \;,\label{3.10}
\end{equation}
and derivatives with respect to the parameters $a$ and $b$. After
some straightforward calculation, using the results for $A_k$ and
$B_k$ found from Eqs.~(\ref{3.8}) and (\ref{3.9}) as described, it
can be shown that
\begin{eqnarray}
\Omega_r&\simeq&-\frac{1}{8\pi^2\beta^3\omega^2}
\sum_{k=1}^{\infty}\frac{(-1)^k}{k^3\sinh\big(
\frac{2\pi^2k}{\beta\omega}\big)} \Big\lbrace \Big\lbrack
8\pi^4k^2\csch^2\Big( \frac{2\pi^2k}{\beta\omega}\Big)\nonumber\\
&&\qquad+4\pi^2k\beta\omega\coth\Big(
\frac{2\pi^2k}{\beta\omega}\Big)
+2\beta^2\omega^2\nonumber\\
&&\qquad+\pi^2k^2(4\pi^2+\beta^2\omega^2-4\beta^2\mu^2)\Big\rbrack\cos(2\pi
k\mu/\omega)\nonumber\\
&&+4\pi k\beta\mu\Big\lbrack \beta\omega+2\pi^2k\coth\Big(
\frac{2\pi^2k}{\beta\omega}\Big) \Big\rbrack \sin(2\pi
k\mu/\omega) \Big\rbrace \;,\label{3.11}
\end{eqnarray}

The presence of the trigonometric functions in Eq.~(\ref{3.11}) is
responsible for the oscillations that occur in thermodynamic
quantities if we go beyond the leading order continuum
approximation. The presence of the hyperbolic functions in
Eq.~(\ref{3.11}) with argument $\frac{2\pi^2k}{\beta\omega}$ means
that unless $\beta\omega>1$ these oscillations will be suppressed.
The oscillatory behaviour should therefore show up for $T<\omega$ and become
more prominent as $T$ is reduced.

The inclusion of $\Omega_r$ in the expression used for the
thermodynamic potential will lead to corrections to the
thermodynamic behaviour of the system beyond what is found using
the continuum approximation of Sec.~\ref{sec2}. We will look first
at how the chemical potential is affected.

\subsection{\label{sec3.2}Chemical potential}

We can define (from Eq.~(\ref{2.23}) )
\begin{equation}
N_r=-\left.\left( \frac{\partial\Omega_r}{\partial\mu}
\right)\right|_{T,\omega} \;,\label{3.12}
\end{equation}
and then use Eq.~(\ref{3.11}) to obtain
\begin{eqnarray}
N_r&\simeq&\frac{\pi}{4\beta^3\omega^3} \sum_{k=1}^{\infty}
\frac{(-1)^k}{\sinh\big(\frac{2\pi^2 k}{\beta\omega}\big)}
\Big\lbrace\Big\lbrack 4\beta^2\mu^2-4\pi^2-\beta^2\omega^2-
8\pi^2\csch^2\big(\frac{2\pi^2 k}{\beta\omega}\big)
\Big\rbrack \nonumber\\ 
&&\times\sin(2\pi k\mu/\omega)+8\pi\beta\mu \coth\big(\frac{2\pi^2
k}{\beta\omega}\big)\cos(2\pi k\mu/\omega)\Big\rbrace
\;.\label{3.13}
\end{eqnarray}
This will make an oscillatory correction to the contribution $N_0$
for the average number of particles found using the continuum
approximation in Sec.~\ref{sec2}.

The chemical potential may be found by solving
\begin{equation}
N=N_0+N_r\label{3.14}
\end{equation}
for $\mu$, where $N_0$ is given by (see Eq.~(\ref{2.25}) with
$\omega_1=\omega_2=\omega_3=\omega$)
\begin{equation}
N_0\simeq\frac{\mu^3}{6\omega^3}-\frac{\mu}{24\omega} \left(
3-\frac{4\pi^2}{\beta^2\omega^2} \right) \;.\label{3.15}
\end{equation}
Obviously the complicated dependence on $\mu$ in $N_r$ as given in
Eq.~(\ref{3.13}) renders an analytical evaluation of $\mu$
difficult. We can simplify by using the assumption that $\mu$ is
large ( since we have already assumed that $\beta\mu\gg1$ and
$\mu\gg\omega$), and keep only the leading term in $\mu$. This
gives
\begin{equation}
N_r\simeq \frac{\pi\mu^2}{\beta\omega^3} \sum_{k=1}^{\infty}
\frac{(-1)^k\sin(2\pi k\mu/\omega)}{\sinh\big(\frac{2\pi^2
k}{\beta\omega}\big)} \;.\label{3.16}
\end{equation}

The continuum approximation in Eq.~(\ref{3.15}) still gives the
leading contribution to $N$  for large $\mu$; however, $N_r$ in
Eq.~(\ref{3.16}) can be more important than the sub-leading term
(proportional to $\mu$) in Eq.~(\ref{3.15}) if the temperature is
low enough. For $\beta\omega$ of the order of 1 or less, the sum
in Eq.~(\ref{3.16}) is well approximated by simply the first term
and it is clear that the oscillations, although present, will have
a very small amplitude. We would therefore expect that for the
temperature range $T\ge\omega$, Eq.~(\ref{2.26}) or
Eq.~(\ref{2.27}) would provide a good approximation for the
chemical potential. However as the temperature is reduced, so that
$\beta\omega\gg1$, the amplitude of the oscillations coming from
Eq.~(\ref{3.16}) become more pronounced and must be taken into
effect.

It is possible to find an asymptotic expansion for $N_r$ in
Eq.~(\ref{3.16}) valid for $\beta\omega\gg1$ (or $T\ll\omega$).
After some calculation, it can be shown that
\begin{equation}
N_r\simeq\frac{\mu^2}{4\omega^2} \left\lbrace \tanh\left(
\frac{\beta\omega}{4}(2\bar{\mu}-1) \right)
+1-2\bar{\mu}\right\rbrace \;,\label{3.17}
\end{equation}
where
\begin{equation}
\bar{\mu}=\frac{\mu}{\omega}-\left\lbrack\frac{\mu}{\omega}\right\rbrack
\;,\label{3.18}
\end{equation}
with $\lbrack x\rbrack$ denoting the largest integer whose value
is less than or equal to $x$. (Thus $0\le\bar{\mu}<1$ can be
assumed in Eq.~(\ref{3.17}).) Similar asymptotic expansions can be
found for the sub-leading terms in $N_r$ given in
Eq.~(\ref{3.13}); however we will not give them here. It can be
shown by numerically evaluating the sum in Eq.~(\ref{3.16}) and
comparison with the analytical approximation in Eq.~(\ref{3.17})
that Eq.~(\ref{3.17}) does give an accurate result for large
values of $\beta\omega$.

If we take the limit $\beta\omega\rightarrow\infty$ in
Eq.~(\ref{3.17}), the result simplifies further to give
\begin{equation}
N_r\simeq\frac{\mu^2}{2\omega^2}\left(
\left\lbrack\frac{\mu}{\omega}+\frac{1}{2}\right\rbrack
-\frac{\mu}{\omega}\right) \;.\label{3.19}
\end{equation}
Strictly speaking, this further approximation is only valid if
$\left\lbrack\frac{\mu}{\omega}+\frac{1}{2}\right\rbrack$ is not
equal to an integer. In this special case, $N_r$ in
Eq.~(\ref{3.16}) or Eq.~(\ref{3.17}) can be seen to vanish, and we
must look at the sub-leading contributions to $N_r$ that follow
from Eq.~(\ref{3.13}).

If we use Eq.~(\ref{3.19}) for $N_r$ in Eq.~(\ref{3.14}) along
with Eq.~(\ref{3.15}) for $N_0$, it is possible to show that the
solution for $\mu$ is given by the simple expression
\begin{equation}
\frac{\mu}{\omega}\simeq\lbrack(6N)^{1/3}\rbrack+\frac{1}{2}
\;.\label{3.20}
\end{equation}
The presence of the greatest integer function in Eq.~(\ref{3.20})
leads to the step-like behaviour first found in the numerical
studies of \cite{SW}. Our results provide a confirmation of this
behaviour by analytical means. For large values of $N$, our result
in Eq.~(\ref{3.20}) shows that these steps will occur roughly for
$N\simeq\ell^3/6$ where $\ell$ is an integer. This agrees with the
``magic numbers'' found by \cite{SW} that occurred for
$N=\ell(\ell+1)(\ell+2)/6$ if we make $N$, and hence $\ell$, large
enough. We have therefore seen how the step-like behaviour of the
chemical potential comes about in an analytical way, and traced
its origin back to the same type of terms that are responsible for
the \dH effect.

\begin{figure}[ht]
\begin{center}
\leavevmode \epsfxsize=125mm \epsffile{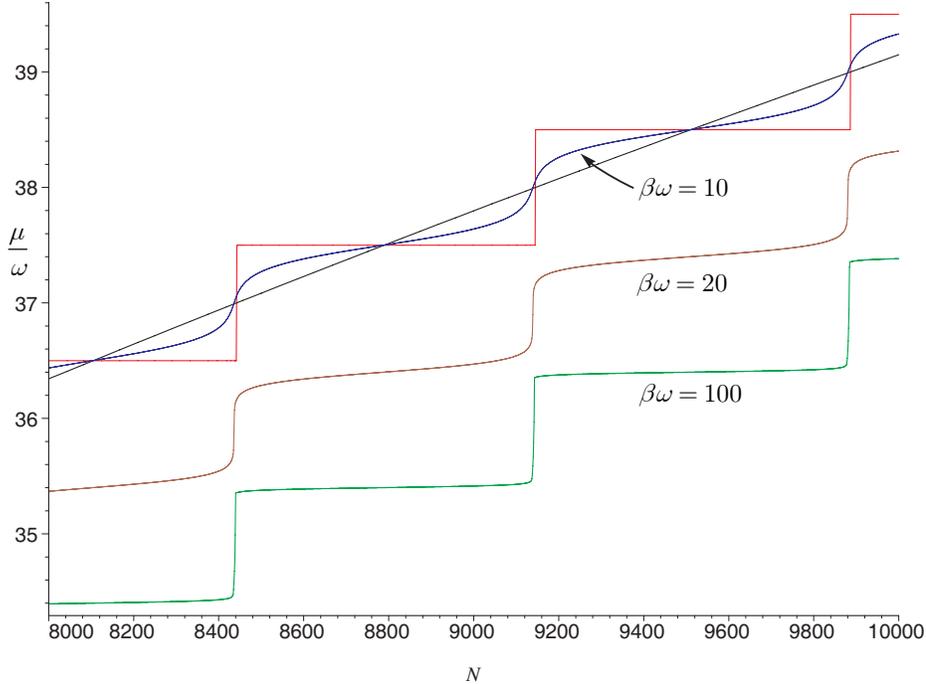}
\end{center}
\caption{(color online) This plot shows $\mu/\omega$ plotted over
a range of $N$ for three sample temperatures. The solid, almost
straight, line shows the result found using the continuum
approximation for the particle number resulting in
$\mu/\omega=(6N)^{1/3}$. The step-function superimposed shows the
result found from the very low temperature approximation of
Eq.~(\ref{3.20}). The smooth sinuous curves give the results for
$T=0.1\omega, T=0.05\omega$ and $T=0.01\omega$. The last two
curves have been displaced from the first one for
clarity.}\label{fig1}
\end{figure}
In Fig.~\ref{fig1} we show the result for $\mu/\omega$ plotted as
a function of $N$ over a range of $N$. We have taken $N$ large,
but the same type of behaviour can be found for smaller values as
well. The continuum approximation for $\mu$ is shown as the smooth
solid, almost straight, line, and the simple analytical
approximation in Eq.~(\ref{3.20}) is superimposed on it. The steps
occur at the magic numbers as predicted. We have displaced the
curves for different values of $\beta\omega$ away from each other for clarity to show
the trend towards the step-like behaviour as the temperature is
reduced. As the temperature is increased, the amplitudes of the
oscillations decreases as mentioned above.

\subsection{\label{sec3.3}Specific heat}

We will first calculate that part of the energy that arises from
$\Omega_r$. Define
\begin{equation}
U_r=\left.\frac{\partial}{\partial\beta}(\beta\Omega_r)
\right|_{\beta\mu,\omega} \;,\label{3.21}
\end{equation}
and use Eq.~(\ref{3.11}) for $\Omega_r$. After some calculation it
can be shown that
\begin{eqnarray}
U_r&\simeq&\frac{\pi\mu^3}{\beta\omega^3}\sum_{k=1}^{\infty}
\frac{(-1)^k}{\sinh\big(\frac{2\pi^2 k}{\beta\omega}\big)}
\sin\left(\frac{2\pi k\mu}{\omega}\right)\nonumber\\
&&\hspace{-1cm}+\frac{3\pi^2\mu^2}{\beta^2\omega^3}\sum_{k=1}^{\infty}
\frac{(-1)^k\cosh\big(\frac{2\pi^2
k}{\beta\omega}\big)}{\sinh^2\big(\frac{2\pi^2
k}{\beta\omega}\big)} \cos\left(\frac{2\pi k\mu}{\omega}\right)
\label{3.22}
\end{eqnarray}
where we only include the two leading order terms in $\mu$ since
it is these two terms we will need to calculate the leading
contribution to the specific heat.

The specific heat was defined in Eq.~(\ref{2.31}). It is
straightforward to show that this expression is equivalent to
\begin{equation}
C=\left.\left(\frac{\partial U}{\partial
T}\right)\right|_{\mu,\omega} - \frac{\left.\left(\frac{\partial
U}{\partial
\mu}\right)\right|_{T,\omega}\left.\left(\frac{\partial
N}{\partial
T}\right)\right|_{\mu,\omega}}{\left.\left(\frac{\partial
N}{\partial \mu}\right)\right|_{T,\omega}} \;,\label{3.23}
\end{equation}
which is more useful for explicitly calculating $C$. The presence
of the second term on the right hand side complicates the
evaluation of $C$, but we will obtain an expansion for $C$ in
powers of $\mu$.

If we look at the first term on the right hand side of
Eq.~(\ref{3.23}) and use $U=U_0+U_r$ with $U_r$ given by
Eq.~(\ref{3.22}) and
\begin{eqnarray}
U_0&=&\left.\frac{\partial}{\partial\beta}(\beta\Omega_0)
\right|_{\beta\mu,\omega}\nonumber\\
&=&\frac{\mu^4}{8\omega^3}-\frac{\mu^2}{16\omega^3}(\omega^2-4\pi^2T^2)\nonumber\\
&&\qquad\qquad+\frac{1}{1920\omega^3}(17\omega^4+40\pi^2\omega^2T^2-112\pi^4T^4)
\label{3.24}
\end{eqnarray}
we obtain
\begin{equation}
\left.\left(\frac{\partial U}{\partial
T}\right)\right|_{\mu,\omega}\simeq\frac{\pi^2\mu^2T}{2\omega^3}+
\left.\left(\frac{\partial U_r}{\partial
T}\right)\right|_{\mu,\omega} \;.\label{3.25}
\end{equation}
From Eq.~(\ref{3.22}), counting powers of $\mu$, it can be
observed that the second term on the right hand side of
Eq.~(\ref{3.25}) will contain expressions in $\mu^3,\mu^2,\ldots$;
thus, it appears as if the leading order behaviour of the specific
heat will be $\mu^3$, rather than $\mu^2$ as predicted by the
continuum approximation. However, we will show that the $\mu^3$
part of Eq.~(\ref{3.25}) cancels with a similar expression coming
from the second term in Eq.~(\ref{3.23}) leaving the overall
leading behaviour of the specific heat as $\mu^2$ rather than
$\mu^3$. In any case, we must work consistently to order $\mu^2$,
and to this order only the two contributions to $U_r$ that we have
written down in Eq.~(\ref{3.22}) are necessary. We will drop all
terms that result in explicit factors of $\mu$ in $C$ that are of
order $\mu$ and lower.

Using Eq.~(\ref{3.24}) we have
\begin{equation}
\left.\left(\frac{\partial U}{\partial
\mu}\right)\right|_{T,\omega}\simeq\frac{\mu^3}{2\omega^3}+
\left.\left(\frac{\partial U_r}{\partial
\mu}\right)\right|_{T,\omega} \;,\label{3.26}
\end{equation}
and it is noted that the second term on the right hand side contains terms
involving $\mu^3,\mu^2,\ldots$. From Eq.~(\ref{3.14}) and
Eq.~(\ref{3.15}) we have
\begin{equation}
\left.\left(\frac{\partial N}{\partial
T}\right)\right|_{\mu,\omega}\simeq\frac{\pi^2\mu T}{3\omega^3}+
\left.\left(\frac{\partial N_r}{\partial
T}\right)\right|_{\mu,\omega} \;,\label{3.27}
\end{equation}
with the second term on the right hand side involving
$\mu^2,\mu,\ldots$, as well as
\begin{equation}
\left.\left(\frac{\partial N}{\partial
\mu}\right)\right|_{T,\omega}\simeq\frac{\mu^2}{2\omega^3}+
\left.\left(\frac{\partial N_r}{\partial
\mu}\right)\right|_{T,\omega} \;,\label{3.28}
\end{equation}
with the second term on the right hand side involving
$\mu^2,\mu,\ldots$. A simple counting of powers of $\mu$, using
the expressions for $U_r$ and $N_r$, shows that the second term on
the right hand side of Eq.~(\ref{3.23}) does begin at order
$\mu^3$.

It now remains to use the explicit results for $U_r$ given in
Eq.~(\ref{3.22}) and $N_r$ given in Eq.~(\ref{3.13}) and evaluate
$C$ for large $\mu$ keeping terms of order $\mu^3$ and $\mu^2$.
After some calculation it can be shown that the order $\mu^3$
terms cancel leaving
\begin{equation}
C\simeq\frac{\pi^2 T\mu^2}{6\omega^3} \Big\lbrace
1+\Sigma_1-12\frac{\Sigma_2^2}{\Sigma_3} \Big\rbrace
\;,\label{3.29}
\end{equation}
with
\begin{eqnarray}
\Sigma_1&=&12\sum_{k=1}^{\infty}(-1)^k\left\lbrace
\frac{\cosh\theta_k}{\sinh^2\theta_k} -\frac{\pi^2 kT}{\omega}
\left(\frac{1}{\sinh\theta_k}+\frac{2}{\sinh^3\theta_k} \right)
\right\rbrace\nonumber\\
&&\qquad\qquad\qquad\times\cos\left(2\pi k\frac{\mu}{\omega}\right) \;,\label{3.30}\\
\Sigma_2&=&\sum_{k=1}^{\infty}(-1)^k\left\lbrace
\frac{1}{\sinh\theta_k} -\frac{2\pi^2 kT}{\omega}
\frac{\cosh\theta_k}{\sinh^2\theta_k} \right\rbrace\sin\left(2\pi
k\frac{\mu}{\omega}\right) \;,\label{3.31}\\
\Sigma_3&=&1+\sum_{k=1}^{\infty}
\frac{(-1)^k4\pi^2k}{\beta\omega\sinh\theta_k}\cos\left(2\pi
k\frac{\mu}{\omega}\right) \;,\label{3.32}
\end{eqnarray}
where
\begin{equation}
\theta_k=\frac{2\pi^2k}{\beta\omega} \label{3.33}
\end{equation}
has been defined to save a bit of writing. The continuum
approximation of Sec.~\ref{sec2} can be regained by dropping all
of the terms in $\Sigma_{1,2,3}$ that have arisen from the \dH
part of the thermodynamic potential. This gives the familiar
linear dependence on temperature~\cite{BR}.

The numerical calculations of \cite{SW} showed that as the
temperature got sufficiently small there was a significant
departure from the linear temperature dependence in the specific
heat. Once again we will show that this follows from our
analytical method, and that the origin of this behaviour is in the
\dH part of the thermodynamic potential. To do this we will
evaluate the asymptotic expansion of the three sums defined in
Eqs.~(\ref{3.30}--\ref{3.32}) for large values of $\beta\omega$.
Leaving out the technical details of this for brevity, we find the
approximate forms
\begin{eqnarray}
\Sigma_1&\simeq&\frac{3\beta^3\omega^3(2\bar{\mu}-1)^2}
{16\pi^2\cosh^2\left(\frac{\beta\omega}{4}(2\bar{\mu}-1) \right)}
-1
\;,\label{3.34}\\
\Sigma_2&\simeq&-\frac{\beta^2\omega^2(2\bar{\mu}-1)}
{16\pi\cosh^2\left(\frac{\beta\omega}{4}(2\bar{\mu}-1) \right)}
\;,\label{3.35}\\
\Sigma_3&\simeq&\frac{\beta\omega}
{4\cosh^2\left(\frac{\beta\omega}{4}(2\bar{\mu}-1) \right)}
\;,\label{3.36}
\end{eqnarray}
with $\bar{\mu}$ defined as in Eq.~(\ref{3.18}). The results in
Eqs.~(\ref{3.34}--\ref{3.36})  can be checked against a numerical
evaluation of the sums defined by Eqs.~(\ref{3.30}--3.32) and
found to be accurate for $\beta\omega\simeq10$ and $\bar{\mu}$ not
too close to 0 or 1. Once $\beta\omega\simeq100$ the results
become very accurate even for $\bar{\mu}$ close to 0 and 1. Thus
for $T\le\omega/100$, the simple expressions in
Eqs.~(\ref{3.34}--\ref{3.36}) become reliable approximations for
$\Sigma_{1,2,3}$.

If we use Eqs.~(\ref{3.34}--\ref{3.36}) in the expression for the
specific heat in Eq.~(\ref{3.29}) the result can be shown to
vanish. The \dH contribution to the specific heat cancels the
continuum approximation to the leading order we are working to.
The specific heat therefore vanishes as $T\rightarrow0$ faster
than $T$. This is completely consistent with the numerical results
found in ~\cite{SW}. Because the \dH approximation, as well as the
asymptotic evaluation of the sums leading to
Eqs.~(\ref{3.34}--\ref{3.36}) neglect terms that are exponentially
suppressed, we suspect that the specific heat vanishes like
$e^{-\alpha\omega/T}$ for some constant $\alpha$ as
$T\rightarrow0$, but we have not been able to establish this in
any simple way. Further support for this belief follows from the
result we are able to establish for the 1-dimensional gas in
Sec.~\ref{sec1D} below.  A more refined estimate of the sums, as
well as the \dH contribution would reveal the exact nature of the
$T\rightarrow0$ limit.

\begin{figure}[htb]
\begin{center}
\leavevmode \epsfxsize=125mm \epsffile{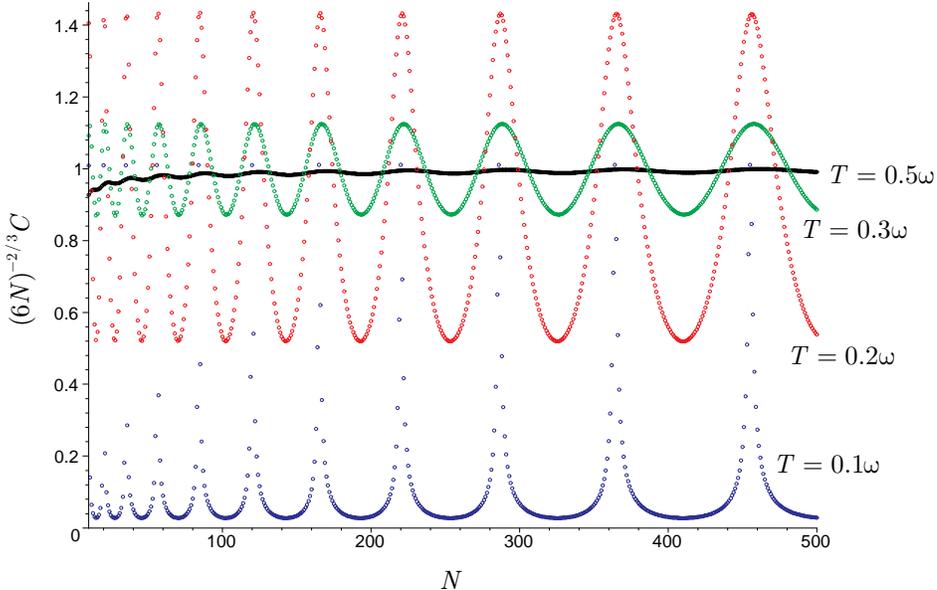}
\end{center}
\caption{(color online) This plot shows $C/(6N)^{2/3}$ plotted
over a range of $N$ for four sample temperatures
$T=0.1\omega,T=0.2\omega,T=0.3\omega$ and $T=0.5\omega$. As the
temperature increases the oscillation amplitude decreases and the
curves approach the continuum limit of 1 in the scaled specific
heat. As the temperature decreases there are significant
deviations from the result found from using the continuum limit,
and for very small temperatures the specific heat starts to become
vanishingly small.}\label{fig2}
\end{figure}

For $T\ge\omega/100$, but still small, the results in
Eqs.~(\ref{3.34}--\ref{3.36}) start to become less reliable. As a
check on our results against the numerical ones of \cite{SW} we
plot the specific heat as found from Eq.~(\ref{3.29}) to
demonstrate the \dH oscillations. This is shown in
Fig.~\ref{fig2}. As the temperature increases we do find, as
expected, that the contribution from $\Sigma_{1,2,3}$ becomes
smaller exponentially, and the linear behaviour with temperature
is regained.

\section{\label{sec4}One and two dimensions}\setcounter{equation}{0}

We will examine the thermodynamics of trapped Fermi gases in one
and two spatial dimensions, since these cases may be of relevance
as limiting cases of the 3-dimensional gas in some situations.
Because the methods used are similar to those described above in
the 3-dimensional case we will be brief here and only exhibit key
results that show a difference with results already obtained.

\subsection{\label{sec1D}One dimension}

With $D=1$, $Z(\beta)$ in Eq.~(\ref{2.15}) has simple poles at
$\beta=\beta_k$ with $\beta_k$ still defined by Eq.~(\ref{3.2}).
It is easy to show that the contribution from the poles with
$k\ne0$ to $\mz(E)$ is
\begin{equation}
\mz_r(E)=-\frac{\omega}{2\pi^2}\sum_{k=1}^{\infty}\frac{(-1)^k}{k^2}
\cos\left( 2\pi k\frac{E}{\omega} \right) \;.\label{4.1}
\end{equation}
The contribution from the pole at $\beta=0$ was given in
Eq.~(\ref{2.18a}). After some calculation, making the
approximation $\mu\gg T$, it can be shown that
\begin{equation}
\Omega_r\simeq\frac{1}{\beta}\sum_{k=1}^{\infty}
\frac{(-1)^k\cos(2\pi k\mu/\omega)}{k\sinh\big(\frac{2\pi^2
k}{\beta\omega} \big)} \;.\label{4.2}
\end{equation}
The calculation is very similar to the $D=3$ case, so details will
not be given here. The continuum approximation for $\Omega$,
called $\Omega_0$, was given in Eq.~(\ref{2.22a}).

Using Eq.~(\ref{2.22a}) and Eq.~(\ref{4.2}) in the general
expression for the average particle number $N$ in Eq.~(\ref{2.23})
results in
\begin{equation}
N\simeq\frac{\mu}{\omega}+\frac{2\pi}{\beta\omega}
\sum_{k=1}^{\infty}\frac{(-1)^k\sin(2\pi
k\mu/\omega)}{\sinh\big(\frac{2\pi^2 k}{\beta\omega} \big)}
\;.\label{4.3}
\end{equation}
The first term on the right hand side comes from $\Omega_0$, and
the second term from $\Omega_r$, so that the continuum
approximation would yield simply
\begin{equation}
\mu\simeq\omega N \label{4.4}
\end{equation}
in this case. Again for large particle numbers we expect
$\mu\gg\omega$ to be a valid approximation. For $T\ll\omega$ we
can again find the asymptotic form of the sum in Eq.~(\ref{4.3})
and obtain
\begin{equation}
N\simeq\left\lbrack\frac{\mu}{\omega}\right\rbrack+\frac{1}{2}
+\frac{1}{2}\tanh\left\lbrace \frac{\beta\omega}{4}\left(
\frac{2\mu}{\omega}-2 \left\lbrack\frac{\mu}{\omega}\right\rbrack
-1\right) \right\rbrace \;,\label{4.5}
\end{equation}
and this proves to be very accurate for low values of $T$.

The internal energy can be found to be
\begin{equation}
U\simeq\frac{\mu^2}{2\omega}+\frac{\pi^2}{6\omega}T^2+\frac{\omega}{24}+U_r
\;,\label{4.6}
\end{equation}
with
\begin{equation}
U_r\simeq\frac{2\pi}{\beta}\sum_{k=1}^{\infty}(-1)^k
\left\lbrace\frac{\mu}{\omega}\,\frac{\sin(2\pi
k\mu/\omega)}{\sinh\theta_k}
+\frac{\pi}{\beta\omega}\,\frac{\cosh\theta_k}{\sinh^2\theta_k}\sin(2\pi
k\mu/\omega) \right\rbrace \label{4.7}
\end{equation}
and $\theta_k$ defined by Eq.~(\ref{3.33}). We can calculate the
specific heat using Eq.~(\ref{3.23}) with the result
\begin{equation}
C\simeq\frac{\pi^2 T}{3\omega} \Big\lbrace
1+\Sigma_1-12\frac{\Sigma_2^2}{\Sigma_3} \Big\rbrace
\;,\label{4.8}
\end{equation}
The sums $\Sigma_{1,2,3}$ are the same as those given in
Eqs.~(\ref{3.30}--\ref{3.33}). The result in Eq.~(\ref{4.8}) is
very similar to that found in Eq.~(\ref{3.29}) in the
3-dimensional case apart from the overall factor that can be
recognized as the continuum approximation for the specific heat of
the 1-dimensional gas. We can therefore conclude immediately that
as $T\rightarrow0$, the specific heat vanishes faster than the
linear temperature dependence deduced from the continuum
approximation.

\begin{figure}[ht]
\begin{center}
\leavevmode \epsfxsize=125mm \epsffile{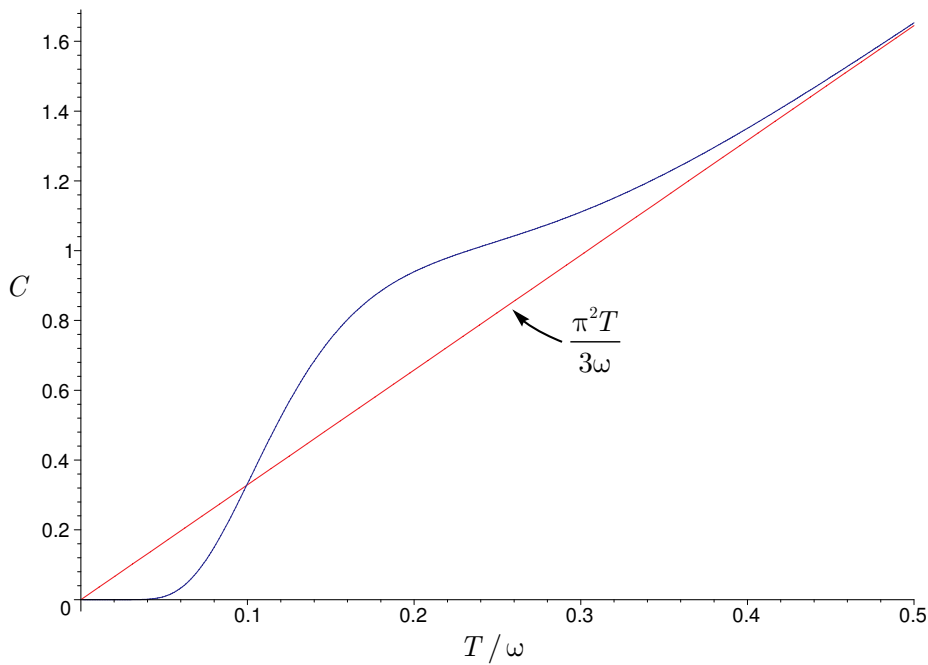}
\end{center}
\caption{(color online) This plot shows the specific heat plotted
over a range of temperature for the 1-dimensional gas. As the
temperature increases the specific heat approach the continuum
limit that behaves linearly on the temperature. As the temperature
decreases the result deviates substantially from the linear result
and the curve decays exponentially fast as found in
Eq.~(\ref{4.9}).}\label{fig3}
\end{figure}
Because of the simplicity of the $D=1$ case, we can shed some
light on the behaviour of the specific heat as $T\rightarrow0$. We
note that the continuum approximation for the specific heat in
Eq.~(\ref{4.4}) is also a solution when the \dH contributions to
$N$ are included, since the $\sin$ term vanishes for integral $N$.
The sum $\Sigma_2$ defined by Eq.~(\ref{3.31}) vanishes as well if
$\mu=\omega N$, and we can use the approximation for $\Sigma_1$
given by Eq.~(\ref{3.34}), valid for $T\ll\omega$ to obtain from
Eq.~(\ref{4.8}) the simple result
\begin{equation}
C\simeq\frac{\omega^2}{16T^2}\exp\left(-\frac{\omega}{2T} \right)
\;.\label{4.9}
\end{equation}
This demonstrates that the specific heat does vanish exponentially
fast as $T\rightarrow0$, not linearly. The role of the \dH part of
the thermodynamic potential is again responsible for this
behaviour. As a check on this conclusion we have plotted the
specific heat as a function of temperature in Fig.~\ref{fig3}. The
results can be seen to be consistent with our analytical result in
Eq.~(\ref{4.9}) as $T\rightarrow0$. For larger values of the
temperature the specific heat approaches the linear temperature
dependence predicted by the continuum approximation. This can also
be seen from the expression obtained in Eq.~(\ref{4.8}) since as
$\beta\omega$ becomes small, the sums $\Sigma_{1,2,3}$ start to
vanish as a consequence of the hyperbolic functions present in the
expressions.

\subsection{\label{sec2D}Two dimensions}

Using $D=2$ in Eq.~(\ref{2.15}) we find
\begin{equation}
\mz_r(E)=-\frac{1}{2\pi^2} \sum_{k=1}^{\infty}\frac{1}{k^2}
\Big\lbrace E\cos(2\pi k E/\omega)-\frac{\omega}{\pi k} \sin(2\pi
k E/\omega) \Big\rbrace \;.\label{4.10}
\end{equation}
This results in
\begin{eqnarray}
\Omega_r&\simeq&\frac{1}{2\pi\beta}\sum_{k=1}^{\infty} \frac{1}{k^2}
\left\lbrace 2\pi k\,\frac{\mu}{\omega}\,\frac{\cos(2\pi k
E/\omega)}{\sinh\theta_k}\right.\nonumber\\
&&\qquad\qquad\qquad\left. -\left\lbrack
\frac{1}{\sinh\theta_k}+\theta_k
\frac{\cosh\theta_k}{\sinh^2\theta_k} \right\rbrack\sin(2\pi k
E/\omega)\right\rbrace \;,\label{4.11}
\end{eqnarray}
with $\theta_k$ defined by Eq.~(\ref{3.33}) if we make the same
approximations as in the $D=3$ case. The $(-1)^k$ factor in the
summand, present for $D=1,3$, is absent here, a result that is
true for any even dimension.

The average particle number can be shown to be
\begin{equation}
N\simeq\frac{\mu^2}{2\omega^2}-\frac{1}{12}+\frac{\pi^2}{6\beta^2\omega^2}+N_r
\;,\label{4.12}
\end{equation}
with
\begin{equation}
N_r\simeq\frac{2\pi}{\beta\omega}\sum_{k=1}^{\infty} \Big\lbrace
\frac{\mu}{\omega}\,\frac{\sin(2\pi k \mu/\omega)}{\sinh\theta_k}
+ \frac{\pi}{\beta\omega}\,\frac{\cosh\theta_k}{\sinh^2\theta_k}
\cos(2\pi k \mu/\omega)\Big\rbrace \label{4.13}
\end{equation}
the contribution coming from the \dH part of $\Omega$.

In the continuum approximation (dropping the term in $N_r$) we
find
\begin{equation}
\frac{\mu}{\omega}\simeq(2N)^{1/2}\;,\label{4.14}
\end{equation}
to leading order for large $N$. However the \dH part of the
particle number can be shown to lead to the step-like behaviour we
saw previously in the 3-dimensional case. In the low temperature
limit ($T\ll\omega$) an asymptotic analysis of Eq.~(\ref{4.13})
can be used to show that
\begin{equation}
N\simeq\frac{1}{2} \left\lbrack\frac{\mu}{\omega}\right\rbrack^2
-\frac{3}{2}
\left\lbrack\frac{\mu}{\omega}\right\rbrack+2\left\lbrack\frac{\mu}{\omega}\right\rbrack
\tanh\left\lbrace\frac{\beta\omega}{2}\left(\frac{\mu}{\omega} -
\left\lbrack\frac{\mu}{\omega}\right\rbrack \right)\right\rbrace
\;.\label{4.15}
\end{equation}
Solving this expression for $\mu$ gives rise to the approximate
analytical solution
\begin{equation}
\frac{\mu}{\omega}\simeq
\left\lbrack(2N)^{1/2}+\frac{1}{2}\right\rbrack \;.\label{4.16}
\end{equation}
(Again we remind the reader that the square brackets in
Eqs.~(\ref{4.15}) and (\ref{4.16}) denote the greatest integer
function.) We therefore conclude that the step-like behaviour
found earlier for the chemical potential when $D=3$ is also found
for the 2-dimensional gas. We have checked this result by solving
Eq.~(\ref{4.12}) numerically and found that the approximation in
Eq.~(\ref{4.16}) becomes increasingly accurate as $T$ is reduced.
Because the results resemble the similar behaviour exhibited in
Fig.~\ref{fig1} we will not show them here. The jumps in the
chemical potential occur, for large $N$, when
$N\simeq\frac{1}{2}\ell^2$ for integral $\ell$. (The exact result
is $\frac{1}{2}\ell(\ell+1)$ analogously to the 3-dimensional case
studied in Ref.~\cite{SW}.)

It is straightforward to obtain expressions for the internal
energy and specific heat. Once again the results are similar to
those found in the 3-dimensional case, this time shown in
Fig.~\ref{fig2}, so need not be exhibited in detail. As
$T\rightarrow0$ the result for the specific heat vanishes much
faster than the linear approximation
$C\simeq{\pi^2T}(2N)^{1/2}/(3\omega)$ predicted by the continuum
approximation.

\section{\label{sec2anis}Anisotropic 3-dimensional potential}\setcounter{equation}{0}
\subsection{\label{sec2an}Thermodynamic potential}

As in the isotropic case, we will apply the analysis used by
Sondheimer and Wilson~\cite{Sond} for the \dH effect to obtain the
thermodynamic potential. This allows a clear separation of the
terms responsible for the continuum approximation and those that
lead to oscillatory behaviour in thermodynamic expressions. The
calculation of the thermodynamic potential will be used in the
next two sections to obtain the chemical potential, particle
number, and the specific heat.

Because we have already outlined the details of the
Sondheimer-Wilson~\cite{Sond} method above, we will begin with the
main result that expresses the thermodynamic potential as (see
(\ref{2.12}))
\begin{equation}
\Omega=-\frac{1}{4}\beta\int_{0}^{\infty}dE\,\frac{\mz(E)}{\cosh^2\frac{1}{2}\beta(E-\mu)}\label{2an.1}
\end{equation}
where $\mz(E)$ is related to the inverse Laplace transform of the
partition function $Z(\beta)$ for the gas with $\mu=0$. (See
(\ref{2.5}) and (\ref{2.8}).)

To evaluate $\mz(E)$ in Eq.~(\ref{2.8}) we require the partition
function. For the case of a free Fermi gas confined by a simple
harmonic oscillator potential we have the result in Eq.~(\ref{2.15}) with $D=3$ taken there. The heart of
the calculation consists of using Eq.~(\ref{2.15}) in
Eq.~(\ref{2.8}), closing the contour in the left-hand side of
the complex plane, and evaluating the result by residues as in the isotropic case considered in Sec.~\ref{sec3}.

As explained above, and as should be evident from
Eq.~(\ref{2.15}), the poles of the integrand of Eq.~(\ref{2.8})
depend on the ratios of the three oscillator frequencies. There
will be a pole of order 5 at $\beta=0$ as well as poles along the
imaginary axis. It is easy to see from Eq.~(\ref{2.15}) that
there are poles away from the origin at $\beta=\beta_j$ where
\begin{equation}
\beta_j=\frac{2\pi ik_j}{\omega_j}\label{2an.5}
\end{equation}
with $j=1,2,3$ and $k_j=0,\pm1,\pm2,\ldots$. The ratio of
oscillator frequencies determines the order of these poles, and
the order affects the residue. If none of the oscillator
frequencies are rational multiples of each other, then the poles
for $\beta_{1,2,3}$ will all be distinct and therefore simple.
However if any of the trap frequencies are rational multiples of
any of the others, then some of the poles along the imaginary axis
will no longer be simple. Therefore a complete analysis of the
problem requires investigating a variety of different cases
depending on the relative ratios of the three oscillator
frequencies. In order to avoid too many different cases, and to
stop the paper being overly lengthy, we will only look at the case
of an axisymmetric trap characterized by
\begin{equation}
\omega_1=\omega_2=\omega\;,\omega_3=\lambda\omega\;,\label{2an.6}
\end{equation}
where $\lambda$ is some real number that characterizes the
anisotropy. There are then only two distinct cases to consider:
$\lambda$ rational, and $\lambda$ irrational. The isotropic
results can be recovered in the special case $\lambda=1$.

Regardless of whether $\lambda$ is rational or irrational, there
is a pole of order 5 at $\beta=0$ for the integrand of
Eq.~(\ref{2.8}). If we call $\Omega_0$ the contribution to the
thermodynamic potential that arises from this pole, then
\begin{eqnarray}
\Omega_0&\simeq&-\frac{\mu^4}{24\lambda\omega^3}
+\frac{\mu^2}{48\lambda\omega^3}\Big\lbrack(2+\lambda^2)\omega^2-4\pi^2
T^2\Big\rbrack\nonumber\\
&&-\frac{1}{5760\lambda\omega^3 } \Big\lbrack3(8+3\lambda^4
)\omega^4-80\pi^2T^2\omega^2
(2+\lambda^2)+112\pi^4T^4\Big\rbrack\;,\label{2an.7}
\end{eqnarray}
as follows from the result in (\ref{2.22c}) with (\ref{2an.6})
used. The first term on the right-hand side of Eq.~(\ref{2an.7})
can be recognized as the continuum (or semi-classical)
approximation for the thermodynamic potential that arises from
approximating the sums over the discrete energy levels with
integrals, and is equivalent to results given in Ref.~\cite{BR}.
The other two terms are the next order corrections to the leading
order behaviour. As we will discuss below, even in situations
where the continuum approximation is valid, the leading order term
may not be accurate if the trap becomes too anisotropic. If
$\lambda$ becomes large, it is possible for the next to leading
order terms, as counted in terms of $\mu$, to become
non-negligible.

\subsubsection{\label{sec2an.1}$\lambda$ irrational}

When $\lambda$ is irrational it is easy to see from
Eq.~(\ref{2.15}) that the integrand of Eq.~(\ref{2.8}) has
double poles at
\begin{equation}
\beta=\beta_k=\frac{2\pi i k}{\omega}\label{2an.8}
\end{equation}
and simple poles at
\begin{equation}
\beta=\tilde{\beta}_k=\frac{2\pi i k}{\lambda\omega}\label{2an.9}
\end{equation}
where $k=\pm1,\pm2,\ldots$ in both cases. (Because $\lambda$ is
assumed irrational here, the $\tilde{\beta}_k$ poles are all
distinct from the $\beta_k$ poles.) It is a straightforward
calculation to show that these poles make a contribution to
$\mz(E)$, that we will call $\mz_r(E)$, of
\begin{eqnarray}
\mz_r(E)&=&-\frac{1}{4\pi^2}\sum_{k=1}^{\infty}
\frac{1}{k^2\sin(\pi k\lambda)}\Big\lbrace E\sin\left(2\pi
k\frac{E}{\omega} \right)\nonumber\\
&&+\Big\lbrack\frac{\omega}{\pi k}+\frac{\lambda\omega}{2}\cot(\pi
k\lambda)\Big\rbrack\cos\left(2\pi k\frac{E}{\omega}
\right)\Big\rbrace\nonumber\\
&&+\frac{\lambda\omega}{8\pi^2}\sum_{k=1}^{\infty}
\frac{\cos\left(\frac{2\pi kE}{\lambda\omega}-\pi k
\right)}{k^2\sin^2(\pi k/\lambda)} \;.\label{2an.10}
\end{eqnarray}
This can be used in Eq.~(\ref{2an.1}), and again making the
approximation $T\ll\mu$ we find
\begin{eqnarray}
\Omega_r&\simeq&\frac{1}{4\pi\beta}\sum_{k=1}^{\infty}\frac{1}{k^2\sin(\pi
k\lambda)\sinh\theta_k}\Big\lbrace\Big\lbrack 1+\pi\lambda\cot(\pi
k\lambda)+\theta_k\coth\theta_k\Big\rbrack
\nonumber\\
&&\qquad\qquad\times\cos\left(2\pi k\frac{\mu}{\omega}\right)+2\pi k\frac{\mu}{\omega}\sin\left(2\pi
k\frac{\mu}{\omega}\right)\Big\rbrace\nonumber\\
&&-\frac{1}{4\beta}\sum_{k=1}^{\infty} \frac{\cos\left(\frac{2\pi
k\mu}{\lambda\omega}-\pi k\right)}{k\sin^2(\pi
k/\lambda)\sinh(\theta_k/\lambda)}\;,\label{2an.11}
\end{eqnarray}
where
\begin{equation}
\theta_k=\frac{2\pi^2 k}{\beta\omega}\;.\label{2an.12}
\end{equation}

At this stage we have not made any assumptions about the magnitude
of $\beta\omega$ in relation to $\beta\mu$. A simple observation
that can be made here is that for small values of $\beta\omega$
the presence of the hyperbolic functions in Eq.~(\ref{2an.11})
tends to suppress the contribution of $\Omega_r$ to $\Omega$. Thus
the oscillatory part of $\Omega$ would be expected to become less
important as the temperature is increased (say $T\sim\omega$), and
more important at lower temperatures ($T<\omega$). For values of
$T$ larger than $\omega$ the leading part of $\Omega$ would be
expected to be well described by the continuum approximation.
given by $\Omega_0$. This will be verified in the next two
sections.

\subsubsection{\label{sec2an.2}$\lambda$ rational}

For rational values of $\lambda$ we will define, without any loss
of generality,
\begin{equation}
\lambda=\frac{p}{q}\;,\label{2an.13}
\end{equation}
where $p$ and $q$ are two integers that are relatively prime.
(That is, $p$ and $q$ have no non-trivial common factors.) The
isotropic case considered above is a special case of this,
corresponding to $p=q=1$. The poles of the integrand of
Eq.~(\ref{2.8}) still occur at values of $\beta$ equal to
$\beta_k$ and $\tilde{\beta}_k$ given in Eq.~(\ref{2an.8}) and
Eq.~(\ref{2an.9}), but this time some of the values of $k$ result
in some of the $\tilde{\beta}_k$ poles coinciding with those of
$\beta_k$. It is easy to show that $Z(\beta)$ has triple poles at
$2\pi iqk/\omega$ for $k=\pm1,\pm2,\ldots$; double poles at $2\pi
ik/\omega$ for $k=\pm1,\pm2,\ldots$, but $k\ne\pm q,\pm2q,\ldots$;
simple poles at $2\pi q ki/(p\omega)$ for $k=\pm1,\pm2,\ldots$ but
$k\ne\pm p,\pm2p,\ldots$. After some calculation, assuming that
$\mu\gg T$ as before, it can be shown that
\begin{eqnarray}
\Omega_r&\simeq&\frac{1}{\beta}\sum_{k=1}^{\infty}\Big\lbrace
\Big\lbrack \frac{\mu^2}{2\omega^2pk\sinh(q\theta_k)}
-\frac{p^2+2q^2}{24pq^2k\sinh(q\theta_k)}\nonumber\\
&& -\frac{\pi^2}{2\beta^2\omega^2pk}\Big(
\frac{1}{\sinh(q\theta_k)}+
\frac{2}{\sinh^3(q\theta_k)}\Big)\nonumber\\
&&-\frac{\cosh(q\theta_k)}{2\beta
pqk^2\sinh^2(q\theta_k)}
-\frac{1}{4\pi^2pq^2k^3\sinh(q\theta_k)}\Big\rbrack
\cos( {2\pi qk\mu}/{\omega}-\pi kp)\nonumber\\
&&-\frac{\mu}{2\pi pq\omega k^2}\Big\lbrack
\frac{1}{\sinh(q\theta_k)}+
\frac{q\theta_k\cosh(q\theta_k)}{\sinh^2(q\theta_k)}
\Big\rbrack \sin( {2\pi qk\mu}/{\omega}-\pi kp)\Big\rbrace\nonumber\\
&+&\frac{1}{4\pi\beta}{\sum_{k=1}^{\infty}}^{\prime\prime}\frac{1}{k^2\sin(\pi\lambda
k)\sinh\theta_k}\Big\lbrace
\left\lbrack1+\pi\lambda\cot(\pi\lambda k)+\theta_k\coth\theta_k
\right\rbrack\nonumber\\ 
&&\times\cos(2\pi k\mu/\omega)+\frac{2\pi k\mu}{\omega}\sin(2\pi k\mu/\omega)\Big\rbrace\nonumber\\
&-&\frac{1}{4\beta}{\sum_{k=1}^{\infty}}^{\prime}\frac{\cos\left(\frac{2\pi
k\mu}{\lambda\omega}-\pi k \right)}{k\sin^2(\pi
k/\lambda)\sinh(\theta_k/\lambda)}\;.\label{2an.14}
\end{eqnarray}
Here $\displaystyle{\sum_{k=1}^{\infty}}^{\prime\prime}$ denotes
that terms with $k=q,2q,3q,\ldots$ are omitted from the sum, and
$\displaystyle{\sum_{k=1}^{\infty}}^{\prime}$ denotes that the
terms with $k=p,2p,3p,\ldots$ are omitted. In the isotropic case,
where $p=q=1$, these last two restricted sums both vanish, and we
recover the result for the isotropic case above.

\subsection{\label{sec3an}Particle number and chemical potential}

Having obtained the thermodynamic potential we can use it to
calculate all of the thermodynamic properties of interest. In this
section we will study the chemical potential and its relation to
the particle number. This can be obtained by solving (\ref{2.23})
for $\mu$.

If we split $\Omega=\Omega_0+\Omega_r$ as in (\ref{2.17}) with
$\Omega_0$ given by Eq.~(\ref{2.7}) and $\Omega_r$ given by either
Eq.~(\ref{2.11}) or Eq.~(\ref{2.14}), we can write $N=N_0+N_r$ in
an obvious way to find
\begin{eqnarray}
N_0=\frac{\mu^3}{6\lambda\omega^3}-\frac{\mu}{24\lambda\omega^3}
\left\lbrack(2+\lambda^2)\omega^2-4\pi^2 T^2\right\rbrack
\;.\label{3an.2}
\end{eqnarray}
By temporarily neglecting the contribution $N_r$ to $N$ coming
from $\Omega_r$, we obtain
\begin{equation}
\frac{\mu}{\omega}\simeq(6\lambda
N)^{1/3}+\frac{1}{12}\left(2+\lambda^2-\frac{4\pi^2 T^2}{\omega^2}
 \right)(6\lambda N)^{-1/3}\;,\label{3an.3}
\end{equation}
valid for large values of $\lambda N$. For $\lambda N\gg1$ we have
$\mu\gg\omega$. The result found in Eq.~(\ref{3an.3}) contains the
continuum approximation of Ref.~\cite{BR} and the next to leading
order correction. We will see that the next to leading order term
can have a significant effect for some traps.

The approximation in Eq.~(\ref{3an.3}) has neglected any effects
that arise from the \dH part of the thermodynamic potential. As in
the isotropic case discussed in Sec.~\ref{sec3}, the \dH terms can
make a significant alteration to the continuum result for the
chemical potential for certain temperatures, and lead to the
step-like behaviour discovered in the numerical results of
Ref.~\cite{SW}. Because of the different expressions for $\Omega$
found in the last section depending on the rational or irrational
nature of $\lambda$, we will look at two separate cases.

\subsubsection{\label{sec3an.1}$\lambda$ rational}

With $\lambda$ defined as in Eq.~(\ref{2.13}) we find,
\begin{eqnarray}
N_r&\simeq&\frac{\pi}{\beta\omega}\sum_{k=1}^{\infty} \Big\lbrace
\Big\lbrack \frac{q\mu^2}{p\omega^2\sinh(q\theta_k)}
-\frac{p^2+2q^2}{12pq\sinh(q\theta_k)}\nonumber\\
&&\qquad\qquad +\frac{\pi^2 q}{p\beta^2\omega^2}\Big(
\frac{1}{\sinh(q\theta_k)}+ \frac{2}{\sinh^3(q\theta_k)}\Big)
\Big\rbrack
\sin( {2\pi qk\mu}/{\omega}-\pi kp)\nonumber\\
&&\qquad\qquad+\frac{2\pi q\mu}{ p\beta\omega^2}
\frac{\cosh(q\theta_k)}{\sinh^2(q\theta_k)}
\cos( {2\pi qk\mu}/{\omega}-\pi kp)\Big\rbrace\nonumber\\
&+&\frac{1}{2\beta\omega}{\sum_{k=1}^{\infty}}^{\prime\prime}\frac{1}{k\sin(\pi\lambda
k)\sinh\theta_k}\Big\lbrace \left\lbrack\pi\lambda\cot(\pi\lambda
k)+\theta_k\coth\theta_k
\right\rbrack \sin(2\pi k\mu/\omega)\nonumber\\
&&\qquad\qquad-\frac{2\pi k\mu}{\omega}\cos(2\pi k\mu/\omega)\Big\rbrace\nonumber\\
&-&\frac{1}{2\beta\lambda\omega}{\sum_{k=1}^{\infty}}^{\prime}\frac{\sin\left(\frac{2\pi
k\mu}{\lambda\omega}-\pi k \right)}{\sin^2(\pi
k/\lambda)\sinh(\theta_k/\lambda)}\;.\label{3an.4}
\end{eqnarray}
For $\mu\gg\omega$ the leading contribution would be expected to
be given simply by the first term in the first sum on the right
hand side of Eq.~(\ref{3an.4}). It is the presence of the
trigonometric functions in Eq.~(\ref{3an.4}) that result in the
step-like behaviour of the chemical potential. Because of the
presence of $\sinh(q\theta_k)$ in the summand, where
$\theta_k=2\pi^2 kT/\omega$, it can be seen that the sums converge
very rapidly if $T\sim\omega$. For $T$ close to $\omega$ very good
analytical approximations can be achieved by merely taking the
first term of the sums. Because the temperature dependence of the
sums is dictated by the magnitude of $\exp(-2\pi^2 qT/\omega)$,
the oscillations, although present, would be expected to be
negligible and unobservable for $T$ close to $\omega$. It is only
once $T$ is reduced below $\omega$ that the oscillatory, or
step-like, behaviour will become observable.

For values of $T<\omega$, it is possible to rewrite the sums that
occur in Eq.~(\ref{3an.4}) in a way that is much more rapidly
convergent. (As $T$ becomes smaller than $\omega$, the sums in
Eq.~(\ref{3an.4}) converge much more slowly rendering their
evaluation more difficult. In addition, it is not sufficient to
retain the first term, or even few terms, and achieve a reliable
analytical approximation.) One way to obtain a reliable analytical
approximation for $T$ small in comparison with $\omega$ is to make
use of the Poisson summation formula, familiar from the well-known
Jacobi inversion formula for the $\theta$-function~\cite{WW}. It
can be shown that
\begin{eqnarray}
\sum_{k=1}^{\infty}\frac{\sin(2\pi km-\pi kp)}{\sinh(2\pi^2
bk)}&=&\frac{1}{4\pi b}\tanh\left\lbrack
\frac{1}{2b}(\bar{m}-\bar{p}/2) \right \rbrack-\frac{1}{2\pi
b}(\bar{m}-\bar{p}/2)\nonumber\\
&&\hspace{-4cm}+\frac{\sinh\left\lbrack
\frac{1}{b}(\bar{m}-\bar{p}/2) \right \rbrack}{2\pi
b}\sum_{k=1}^{\infty} \left\lbrack \cosh(k/b)+ \cosh\left(
\frac{1}{b}(\bar{m}-\bar{p}/2) \right) \right\rbrack^{-1}
\;,\label{3an.5}
\end{eqnarray}
where
\begin{eqnarray}
\bar{m}&=&m-\lbrack m\rbrack\;,\label{3an.6}\\
\bar{p}&=&p\;{\rm mod}\,2\;.\label{3an.7}
\end{eqnarray}
Here $\lbrack x\rbrack$ denotes the greatest integer that does not
exceed $x$. The result in Eq.~(\ref{3an.5}) is sufficient to
evaluate the leading term on the right hand side of
Eq.~(\ref{3an.4}). Other sums that occur in Eq.~(\ref{3an.4}) may
be evaluated in a similar manner. The advantage of this
transformation is that for small values of $b$ (where for
Eq.~(\ref{3an.4}) we have $b\propto T/\omega$) the slowly
convergent sum on the left hand side of Eq.~(\ref{3an.5}) is
converted into a rapidly convergent sum on the right hand side. In
fact, for $b\ll1$, we can safely neglect the sum that occurs on
the right hand side of Eq.~(\ref{3an.5}) and obtain a reliable
analytical approximation from the first two terms on the right
hand side, giving a very simple approximation. It is worth noting
that the appearance of $\bar{m}$ and $\bar{p}$ as defined in
Eqs.~(\ref{3an.6}) and (\ref{3an.7}) on the right hand side of
Eq.~(\ref{3an.5}) is necessary to preserve the periodic structure
in $m$ and $p$ found in the original sum on the left hand side; it
is incorrect to simply take $m$ and $p$ on the right hand side of
Eq.~(\ref{3an.5}). An alternate way of obtaining the analytical
terms on the right hand side of Eq.~(\ref{3an.5}) makes use of the
Euler-Maclaurin summation formula. This will not give the last
term on the right hand side of Eq.~(\ref{3an.5}) as it neglects
exponentially small terms.

If we take $b\ll1$, and neglect the sum on the right hand side of
Eq.~(\ref{3an.5}), it is possible to approximate the result in an
even simpler way. It is easy to show that
\begin{equation}
\sum_{k=1}^{\infty}\frac{\sin(2\pi km-\pi kp)}{\sinh(2\pi^2
bk)}\simeq\frac{1}{4\pi b} \left\lbrace
1-\bar{p}+2\left(\left\lbrack m+\bar{p}/2\right\rbrack-m\right)
\right\rbrace\;.\label{3an.8}
\end{equation}
($\lbrack x\rbrack$ is the greatest integer function mentioned
above.) If we keep only the term of order $\mu^2$ on the right
hand side of Eq.~(\ref{3an.4}), that would be expected to dominate
the result for large values of $\mu\gg\omega$, the following
simple approximation can be found:
\begin{equation}
N_r\simeq \frac{\mu^2}{4p\omega^2} \left\lbrace
1-\bar{p}+2\left(\left\lbrack
\frac{q\mu}{\omega}+\frac{\bar{p}}{2}\right\rbrack-\frac{q\mu}{\omega}\right)
\right\rbrace\;.\label{3an.9}
\end{equation}
For the isotropic oscillator ($p=q=1$) this reduces to the result
in Eq.~(\ref{3.19}). When $T<\omega$, but not necessarily such
that $T\ll\omega$, it is more reliable to use the result of
Eq.~(\ref{3an.5}).

\begin{figure}[thb]
\begin{center}
\leavevmode \epsfxsize=125mm
\epsffile{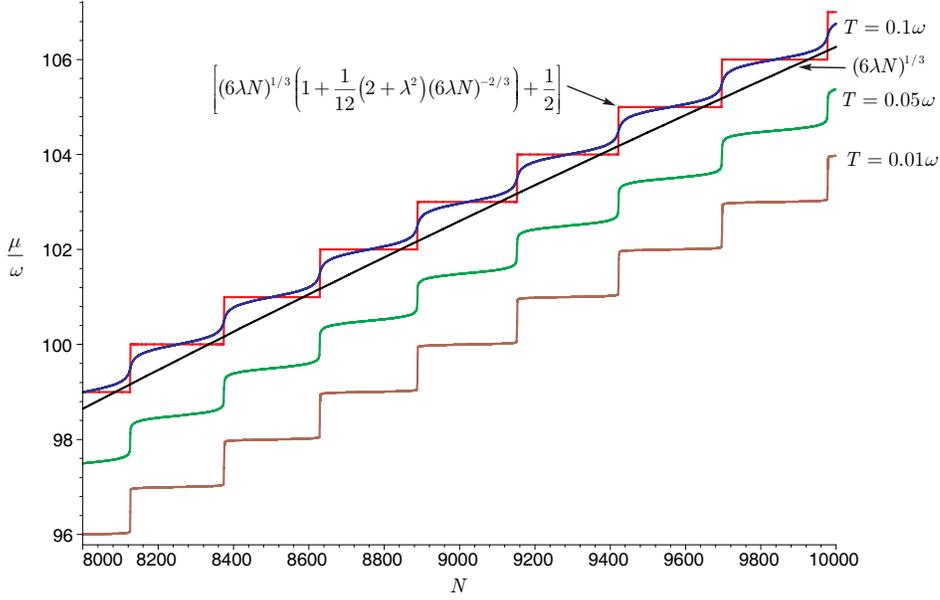}
\end{center}
\caption{(color online) This figure shows the chemical potential
in units of $\omega$ plotted as a function of the particle number
$N$ for a range of temperatures
$T=0.1\omega,0.05\omega,0.01\omega$ with the anisotropy parameter
$\lambda=20$. The curves with $T=0.05\omega$ and $0.01\omega$ have
been displaced vertically by 1.5 and 3 units respectively for
clarity. The almost straight line shows the continuum
approximation $(6\lambda N)^{1/3}$, and the step function over the
$T=0.1\omega$ curve shows the analytical approximation
$\left\lbrack ( 6\lambda N)^{1/3}+\frac{1}{12}(2+\lambda^2)(
6\lambda N)^{-1/3} + \frac{1}{2}\right\rbrack$ found from
Eq.~(\ref{3an.11}) with $p=20,q=1$. (The temperature correction is
negligible here and has been omitted.) }\label{fig1an}
\end{figure}
When the temperature is low enough so that Eq.~(\ref{3an.9}) is a
reliable approximation, it is possible to solve for $\mu$ in terms
of $N$ and obtain the following analytical result:
\begin{equation}
\frac{\mu}{\omega}\simeq\frac{1}{q}\left\lbrace \left\lbrack
\left( 6pq^2N\right)^{1/3}+ \frac{1-\bar{p}}{2}\right\rbrack
+\frac{\bar{p}}{2}\right\rbrace \;.\label{3an.10}
\end{equation}
Because of the greatest integer function in Eq.~(\ref{3an.10}) the
result shows the step-like behaviour found in the numerical
studies of Ref.~\cite{SW}. In obtaining this result we have
approximated $N_r$ with Eq.~(\ref{3an.9}) and only included the
leading order continuum approximation, the order $\mu^3$ term,
from Eq.~(\ref{3an.2}). This means that the result would be
expected to break down in situations where the neglected terms can
become comparable in magnitude to those retained. One case of
direct physical significance occurs when we allow the trap
anisotropy parameter $\lambda$ to become large. In this case the
subleading part of Eq.~(\ref{3an.2}) can lead to a non-trivial
correction to the standard result of $\mu/\omega\simeq(6\lambda
N)^{1/3}$ as obtained in Eq.~(\ref{3an.3}). It is not difficult to
see that a better approximation for the chemical potential, that
takes the \dH part of $\Omega$ into account, is
\begin{eqnarray}
\frac{\mu}{\omega}&\simeq&\frac{1}{q}\Biggl\lbrace \Biggl\lbrack q(
6\lambda
N)^{1/3}\left(1+\frac{1}{12}\left(2+\lambda^2-\frac{4\pi^2T^2}{\omega^2}
\right)( 6\lambda N)^{-2/3} \right)\nonumber\\
&&\qquad\qquad+
\frac{1-\bar{p}}{2}\Biggr\rbrack +\frac{\bar{p}}{2}\Biggr\rbrace
\;,\label{3an.11}
\end{eqnarray}
in place of Eq.~(\ref{3an.10}). (Recall that $\lambda=p/q$ here.)
For small $\lambda$ and $T\ll\omega$ the result in
Eq.~(\ref{3an.10}) is a good approximation. For larger values of
$\lambda$ it is necessary to use the more accurate estimation
Eq.~(\ref{3an.11}).

In Fig.~\ref{fig1an} we show the transition from smooth
oscillations to step-like behaviour of the chemical potential for
a range of $N$ and decreasing temperatures. (Other ranges of $N$
show a similar picture.) The trend towards the step-like behaviour
as the temperature is reduced, predicted by the result in
Eq.~(\ref{3an.11}) is evident. We have chosen a large anisotropy
parameter $\lambda=20$ to illustrate how the leading order
continuum approximation becomes more inaccurate as the anisotropy
increases. Had we plotted the improved expression of
Eq.~(\ref{3an.3}) it would have been centered on the chemical
potential curves; it therefore provides a better estimate for the
chemical potential, although still missing the step-like features
that are present. For smaller values of the anisotropy parameter,
the result $(6\lambda N)^{1/3}$ becomes more accurate, and the
plots look similar to those presented in the isotropic case
described earlier.

\subsubsection{\label{sec3an.2}$\lambda$ irrational}

For irrational values of $\lambda$ we find
\begin{eqnarray}
N_r&\simeq&
\frac{1}{2\beta\omega}{\sum_{k=1}^{\infty}}\frac{1}{k\sin(\pi\lambda
k)\sinh\theta_k}\Big\lbrace \left\lbrack\pi\lambda\cot(\pi\lambda
k)+\theta_k\coth\theta_k
\right\rbrack \sin(2\pi k\mu/\omega)\nonumber\\
&&\qquad\qquad-\frac{2\pi k\mu}{\omega}\cos(2\pi k\mu/\omega)\Big\rbrace\nonumber\\
&-&\frac{1}{2\beta\lambda\omega}{\sum_{k=1}^{\infty}}\frac{\sin\left(\frac{2\pi
k\mu}{\lambda\omega}-\pi k \right)}{\sin^2(\pi
k/\lambda)\sinh(\theta_k/\lambda)}\;.\label{3an.2.1}
\end{eqnarray}
This expression is simply the last two terms of Eq.~(\ref{3an.4})
with the restrictions on the sums removed. For irrational values
of $\lambda$, the confluence of poles in $Z(\beta)$ that gave rise
to the triple pole cannot occur here, and this is why the first
term of Eq.~(\ref{3an.4}) is not present. The leading order
contribution to $N_r$ is proportional to $\mu$ rather than $\mu^2$
as we found in the case of rational $\lambda$ above. This leads to
quantitatively different behaviour for the chemical potential.

It has proven very difficult to obtain reliable asymptotic
approximations for the sums in Eq.~(\ref{3an.2.1}), even for the
leading order part of $N_r$ because of the presence of the
$\sin(\pi\lambda k)$ factor in the summand. The techniques that
prove so useful in the rational case do not work here, as they all
involve the conversion of the sum into an ordinary integral at
some stage, and this requires the integral to converge. The
$\sin(\pi\lambda k)$ factor prevents this here. As a result, our
results are less complete than for the rational case examined
earlier.

Because of the lack of analytical expressions when $\lambda$ is
irrational we will be brief here and only mention the main
conclusions. For values of $\lambda\sim1$ (we used
$\lambda=\surd{3}$) we found that although the oscillations in
$\mu$ were present, they only became noticeable at very low values
of the temperature, typically $T=10^{-2}\omega$. This is expected
as a consequence of the \dH contribution to the particle number
involving $\mu$, rather than $\mu^2$ as in the rational case. In
addition the amplitude of the oscillations is much smaller than in
the rational case. A conclusion is that the continuum
approximation for $\mu$ as given by the first term on the right
hand side of Eq.~\ref{3an.3}, is more reliable, even for
temperatures as low as $T=0.1\omega$, than in the rational case.
For larger values of $\lambda$ (we used $\lambda=\surd{401}$ to
compare with $\lambda=20$ in the rational case) similar
conclusions were found provided that the extra term in the
continuum approximation as given in Eq.~\ref{3an.3} was included.

\subsection{\label{sec4an}Specific heat}

The internal energy $U$ is defined in terms of the thermodynamic
potential $\Omega$ by (\ref{2.24}). The specific heat $C$ is then
defined in terms of $U$ by (\ref{2.31}). With
$\Omega=\Omega_0+\Omega_r$ it can be readily seen that
\begin{eqnarray}
U_0&=&\left. \frac{\partial}{\partial\beta}(\beta\Omega_0)
\right|_{\beta\mu,\omega} \nonumber\\
&\simeq&\frac{\mu^4}{8\lambda\omega^3}
-\frac{\mu^2}{48\lambda\omega^3} \left\lbrack
(2+\lambda^2)\omega^2-12\pi^2T^2\right\rbrack \label{4an.3}
\end{eqnarray}
in agreement with what was found in (\ref{3.24}). As in the
previous section, the result for $U_r$, coming from taking
$\Omega=\Omega_r$ in Eq.~(\ref{2.24}), will be different
depending upon whether $\lambda$ is rational or irrational. The
results for $U_r$ will be given explicitly in Secs.~\ref{sec4an.1}
and \ref{sec4an.2} below. 

Although the internal energy $U$ has a simple representation as
$U=U_0+U_r$, the same is not true for the specific heat because of
$N$ being held fixed in Eq.~(\ref{2.31}). It is easy to see that
Eq.~(\ref{2.31}) can be expressed as in (\ref{3.23}). The second
term on the right hand side of (\ref{3.23}) frustrates any attempt
to express $C$ simply as $C=C_0+C_r$ with $C_r$ the correction to
the continuum approximation coming from the \dH part of $\Omega$.
In order to see the effects that the \dH contributions make to the
specific heat we will initially ignore them, and work out just the
continuum approximation.

We will define $C_0$ to be the continuum approximation for the
specific heat where we take $U=U_0$ and $N=N_0$ in
Eq.~(\ref{3.23}). Using Eqs.~(\ref{4an.3}) and (\ref{3.2}) we
find (for $T\ll\omega$)
\begin{equation}
C_0\simeq\frac{\pi^2\mu^2 T}{6\lambda\omega^3}\;.\label{4an.5}
\end{equation}
The conclusion based on the continuum approximation is that at low
temperatures the specific heat varies linearly with the
temperature. However, as in the isotropic case, and the numerical
studies of Ref.~\cite{SW}, this behaviour turns out to be
significantly altered once the \dH part of $\Omega$ is included.

\subsubsection{\label{sec4an.1}$\lambda$ rational}

If we concentrate on the leading order contribution to the
specific heat for large $\mu$, it would be expected, based on
Eq.~(\ref{4an.5}), that we would find $C_r\propto\mu^2$. We will
therefore work consistently to order $\mu^2$ when we evaluate
$C_r$. If we use $\Omega_r$ as given in Eq.~(\ref{2.14}) for
$\Omega$ in the definition Eq.~(\ref{2.24}) we obtain
\begin{eqnarray}
U_r&\simeq&\sum_{k=1}^{\infty}\Big\lbrace \frac{\pi
q\mu^3}{p\beta\omega^3\sinh(q\theta_k)} \sin( {2\pi
qk\mu}/{\omega}-\pi kp)\nonumber\\
&&+\frac{3 \pi^2
q\mu^2}{p\beta^2\omega^3}\frac{\cosh(q\theta_k)}{\sinh^2(q\theta_k)}
\cos( {2\pi
qk\mu}/{\omega}-\pi kp)\Big\rbrace\nonumber\\
&-&{\sum_{k=1}^{\infty}}^{\prime\prime}\frac{\pi\mu^2}{\beta\omega^2}\frac{\cos(2\pi
k\mu/\omega)}{\sin(\pi\lambda
k)\sinh\theta_k}+\cdots\;.\label{4an.6}
\end{eqnarray}
Terms of order $\mu$ or less have not been kept in this
expression. (Recall that ${}^{\prime\prime}$ on the second sum
means that the terms with $k=q,2q,\ldots$ are omitted and that
$\lambda=p/q$ here. $\theta_k$ was defined in Eq.~(\ref{2.12}).)

When we compute $\displaystyle{\left.\left(\frac{\partial
U_r}{\partial T}\right)\right|_{\mu,\omega}}$, that is needed to
find the complete expression for $C$, it can be seen that there
will be a term of order $\mu^3$ present rather than $\mu^2$ as
expected; however, it turns out that the term of order $\mu^3$
coming from the first term on the right hand side of
Eq.~(\ref{3.23}) cancels with an identical contribution coming
from the second term, leaving $C\propto\mu^2$ as predicted. After
some calculation, assuming that $\mu$ is large, it can be shown
that
\begin{equation}
C\simeq\frac{\pi^2 T\mu^2}{6\lambda\omega^3} \Big\lbrace
1+\Sigma_1-12\frac{\Sigma_2^2}{\Sigma_3} \Big\rbrace
\;,\label{4an.7}
\end{equation}
with
\begin{eqnarray}
\Sigma_1&=&12\sum_{k=1}^{\infty}\left\lbrace
\frac{\cosh(q\theta_k)}{\sinh^2(q\theta_k)} -\frac{\pi^2
 q k}{\beta\omega} \left(\frac{1}{\sinh(q\theta_k)}+\frac{2}{\sinh^3(q\theta_k)}
\right)
\right\rbrace\nonumber\\
&&\qquad\qquad\times\cos\left(\frac{2\pi q k\mu}{\omega}-\pi kp\right),\label{4an.8}\\
\Sigma_2&=&\sum_{k=1}^{\infty}\left\lbrace
\frac{1}{\sinh(q\theta_k)} -q\theta_k
\frac{\cosh(q\theta_k)}{\sinh^2(q\theta_k)}
\right\rbrace\sin\left(\frac{2\pi q k\mu}{\omega}-\pi kp\right),\label{4an.9}\\
\Sigma_3&=&1+\frac{4\pi^2q}{\beta\omega}\sum_{k=1}^{\infty}
\frac{k}{\sinh(q\theta_k)}\cos\left(\frac{2\pi q k\mu}{\omega}-\pi
kp\right) \;.\label{2.240}
\end{eqnarray}
It is noteworthy that the dependence on the restricted summation
in Eq.~(\ref{4an.6}) that could have occurred in the end result
has cancelled. The continuum approximation of Eq.~(\ref{4an.5})
is regained if we set $\Sigma_{1},\Sigma_2$ and $\Sigma_3$ to
zero; this is expected because this is equivalent to dropping the
contribution from the \dH part of $\Omega$.

For very low temperatures it is possible to obtain reliable
analytic approximations for $\Sigma_{1,2,3}$. One way to do this
is to use the Poisson summation formula as we did in
Sec.~\ref{sec3.1} to approximate $N_r$. If we assume that
$qT\ll\omega$, it is possible to derive the following
approximations:
\begin{eqnarray}
1+\Sigma_1&\simeq&\frac{3\beta^3\omega^3\left(m-\frac{1}{q}\lbrack
qm\rbrack-\frac{\bar{p}}{2q} \right)^2} {4\pi^2
q\cosh^2\left(\frac{\beta\omega}{2}\left(m-\frac{1}{q}\lbrack
qm\rbrack-\frac{\bar{p}}{2q} \right) \right)}
\;,\label{2.241}\\
\Sigma_2&\simeq&-\frac{\beta^2\omega^2\left(m-\frac{1}{q}\lbrack
qm\rbrack-\frac{\bar{p}}{2q} \right)} {8\pi
q\cosh^2\left(\frac{\beta\omega}{2}\left(m-\frac{1}{q}\lbrack
qm\rbrack-\frac{\bar{p}}{2q} \right) \right)}
\;,\label{2.242}\\
\Sigma_3&\simeq&\frac{\beta\omega}
{4q\cosh^2\left(\frac{\beta\omega}{2}\left(m-\frac{1}{q}\lbrack
qm\rbrack-\frac{\bar{p}}{2q} \right) \right)} \;.,\label{2.243}
\end{eqnarray}
As before, $\lbrack x\rbrack$ is the greatest integer function and
$\bar{p}=p\;{\rm mod}\,2$. We have not written out the terms in
Eqs.~(\ref{2.241}--\ref{2.243}) that behave like
$\exp(-\alpha/T)$ for constant $\alpha$ as $T\rightarrow0$. It is
now possible to see that as $T$ becomes much smaller than
$\omega/q$ the specific heat behaves like $T\exp(-\alpha/T)$,
demonstrating analytically that there is a deviation from the
linear dependence predicted by the continuum approximation. As $T$
increases, the terms in $\Sigma_{1,2,3}$ in Eq.~(\ref{4an.7})
begin to decrease in importance and the linear behaviour starts to
become a good approximation.

\begin{figure}[thb]
\begin{center}
\leavevmode \epsfxsize=125mm
\epsffile{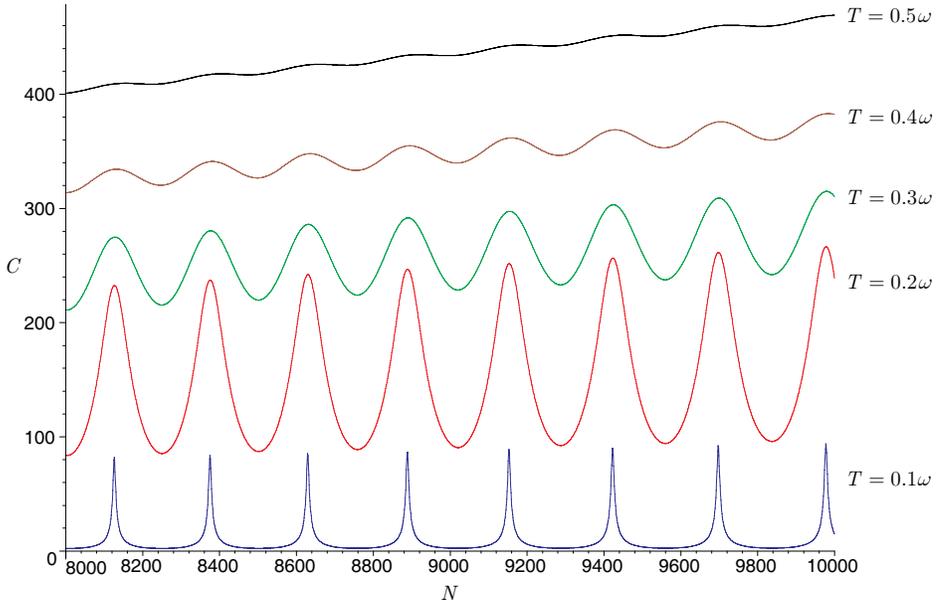}
\end{center}
\caption{(color online) This figure shows the specific heat
plotted as a function of the particle number $N$ over a range of
$N$. The different curves show the results for five different
temperatures. The anisotropy parameter $\lambda=20$. }\label{fig2an}
\end{figure}

\begin{figure}[thb]
\begin{center}
\leavevmode \epsfxsize=125mm
\epsffile{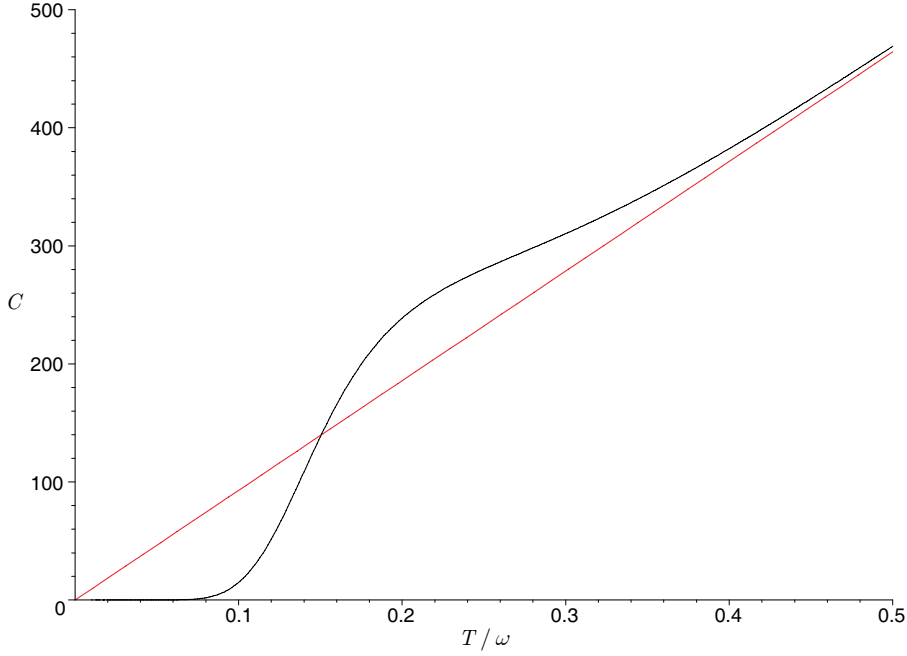}
\end{center}
\caption{(color online) This figure shows the specific heat
plotted as a function of the temperature for $\lambda=20$ and
$N=10^5$ both fixed. The straight line indicates the linear
temperature dependence given by the continuum approximation
$\displaystyle{C=\frac{\pi^2N^{2/3}T}{(6\lambda)^{1/3}\omega}}$.
The other curve shows the effect if the \dH contribution is
included.}\label{fig3an}
\end{figure}

The behaviour of the specific heat with particle number and with
temperature are shown in Figs.~\ref{fig2an} and \ref{fig3an}
respectively. In Fig.~\ref{fig2an} the oscillations in the specific
heat are clearly visible. The amplitude of the oscillations starts
to decrease very rapidly as the temperature is increased. Once $T$
becomes greater than about $0.5\omega$ they all but disappear. As
the temperature is reduced there is the competing effect that the
specific heat starts to vanish exponentially fast as discussed
above and illustrated in Fig.~\ref{fig2an}. Thus there is really a
narrow range of temperature when both the specific heat and the
amplitude of the oscillations are large enough to be potentially
observable. In Fig.~\ref{fig2an} it can be noted that once $T$ is
about $0.1\omega$ there is a dramatic fall-off in the specific
heat from the expected linear behaviour that is fully consistent
with the analytical expressions obtained above.

\begin{figure}[thb]
\begin{center}
\leavevmode \epsfxsize=125mm
\epsffile{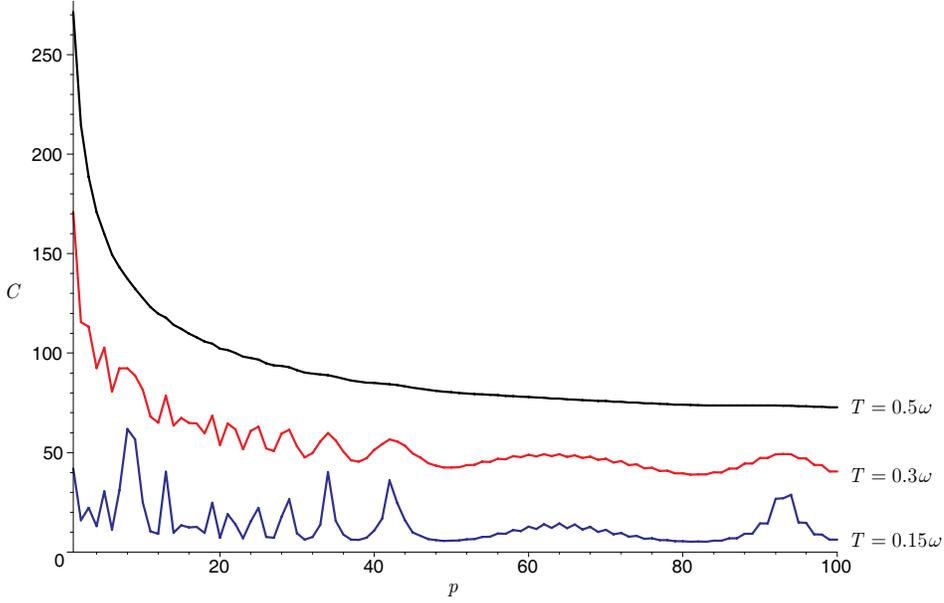}
\end{center}
\caption{(color online) This figure shows the specific heat
plotted as a function of the anisotropy parameter with $N=10^3$
held fixed. $\lambda=p$ here with $p$ an integer. The result is
shown for three different temperatures $T=0.15\omega,0.3\omega$
and $0.5\omega$. }\label{fig4}
\end{figure}

\subsubsection{\label{sec4an.2}$\lambda$ irrational}

As in the calculation of the chemical potential, we were less
successful at obtaining simple analytical expressions here than in
the case of rational $\lambda$. Using Eq.~(\ref{2an.11}) for
$\Omega_r$ in Eq.~(\ref{2.24}) we find
\begin{eqnarray}
U_r&\simeq&\sum_{k=1}^{\infty}\frac{1}{k\sin(\pi k\lambda)}
\Big\lbrace\Big\lbrack \frac{2\pi^3
k}{\beta^3\omega^2\sinh^3\theta_k} +\frac{\pi^3 k}{\beta^3\omega^2\sinh\theta_k}\nonumber\\
&&\qquad\qquad+\frac{\pi^2\lambda\cot(\pi
k\lambda)\cosh\theta_k}{2\beta^2\omega\sinh^2\theta_k}- \frac{\pi
k\mu^2}{\beta\omega^2\sinh\theta_k}\Big\rbrack\cos(2\pi
k\mu/\omega)\nonumber\\
&&+\Big\lbrack\frac{\pi^2k\mu\cosh\theta_k}{\beta^2\omega^2\sinh^2\theta_k}
+\frac{\mu}{2\beta\omega\sinh\theta_k}\Big(\pi\lambda\cot(\pi
k\lambda)+\frac{2\pi^2k\cosh\theta_k}{\beta\omega\sinh\theta_k}
\Big)\Big\rbrack\nonumber\\
&&\qquad\qquad\times\sin(2\pi
k\mu/\omega)\nonumber\\
&&-\sum_{k=1}^{\infty} \Big\lbrace
\frac{\pi\mu}{2\lambda\beta\omega\sin^2(\pi
k/\lambda)\sinh(\theta_k/\lambda)}
\sin\Big(\frac{2\pi k\mu}{\lambda\omega}-\pi k\Big)\nonumber\\
&&\qquad\qquad
+\frac{\pi^2\cosh(\theta_k/\lambda)}{2\lambda\beta^2\omega\sin^2(\pi
k/\lambda)\sinh^2(\theta_k/\lambda)} \cos\Big(\frac{2\pi
k\mu}{\lambda\omega}-\pi k\Big)\Big\rbrace \;.\label{4an.2.1}
\end{eqnarray}
This expression can be contrasted with that found in the rational
case in Eq.~(\ref{4an.6}). The \dH contribution to the internal
energy begins at order $\mu^2$ rather than $\mu^3$. This has a
direct consequence for the specific heat. If we use
Eq.~(\ref{3.23}) to calculate the specific heat to leading order
we find that to leading order in $\mu$ the specific heat is given
by the continuum expression in Eq.~(\ref{4an.5}). For irrational
values of $\lambda$ the \dH contribution cancel out of the
specific heat to leading order. There is a contribution from the
\dH part of the internal energy at sub-leading order $\mu$;
however we will not give the expression here as it has only a
negligible effect.

\section{\label{sec5}Discussion and conclusions}\setcounter{equation}{0}

The main conclusion of this paper is that the origin of the
step-like features found in the numerical calculations of
Schneider and Wallis~\cite{SW} have the same origin as the
periodicity found in the \dH effect and that they can be approximated by analytical means. The general approach of
Sondheimer and Wilson~\cite{Sond}, so useful in the analysis of
the \dH effect, can be used to great effect here to obtain
analytical results for various thermodynamic quantities. In
addition, it is possible to see that within this approach the
continuum approximation used by Butts and Rokhsar~\cite{BR}
corresponds to the neglect of an infinite set of poles in the
Laplace transform of the partition function. It is also possible
to recover the continuum approximation using the methods described
in the general framework of Ref.~\cite{Kirsten}.

In Sec.~\ref{sec3} we obtained approximate analytical results for
the 3-dimensional, isotropic harmonic oscillator potential. In the
case $T\ll\omega$, a very simple result (Eq.~(\ref{3.20})) was
found for the chemical potential that clearly exhibited the
step-like features found in Ref.~\cite{SW}. This was used to
evaluate the specific heat, and it was shown (in agreement with
Ref.~\cite{SW}) that there were significant deviations from the
linear temperature dependence predicted by the continuum
approximation. The specific heat was seen to exhibit a periodic
structure in the particle number.

In Sec.~\ref{sec4} we studied the trapped Fermi gas in 1 and 2
spatial dimensions. For the 1-dimensional gas we were able to show
that as $T\rightarrow0$ the specific heat vanished exponentially
fast in the inverse temperature (Eq.~(\ref{4.9})). The
2-dimensional gas was similar in many ways to the 3-dimensional
case of Sec.~\ref{sec3}.

The same methods that were applied in the isotropic case can be used to examine anisotropic potentials. The technical
details are more involved due to the structure of the poles in the
inverse Laplace transform of the partition function as described
above. In addition to a periodicity in the particle number, there
can also be a periodicity when the trapping potential is altered.
This latter effect was not found for the isotropic case and it is
worth commenting on why this occurs, in contrast with what might
be expected from the \dH effect where the thermodynamics shows a
periodicity in the inverse magnetic field strength. The difference
between the two cases is related to the relationship between the
chemical potential and the particle number. In both cases (\dH and
trapped Fermi gas) the periodic structure results from a
trigonometric dependence on $\mu/\omega$. (For the \dH effect
$\omega$ is the cyclotron frequency associated with the magnetic
field strength.) In the \dH effect, when $\mu$ is solved for in
terms of the particle number the leading order contribution turns
out to be independent of $\omega$. Thus the trigonometric
functions that involve $\mu/\omega$ exhibit the familiar \dH
oscillations as the magnetic field is varied. For the trapped
Fermi gas in an isotropic potential we found $\mu\propto\omega$,
so that $\mu/\omega$ is independent of $\omega$, although there is
a dependence on the particle number. We claim that this is an
artifact of the simplicity of the isotropic potential, and that
the situation for anisotropic potentials is more interesting. 

The treatment presented here has only been performed for the free
Fermi gas. As already mentioned in the introduction, the neglect
of interactions for a single component gas is a good
approximation~\cite{BR,SW}. In Ref.~\cite{BB} a two component
model was studied that allowed for an interaction between the two
spin components. A numerical study showed that as the interaction
strength is increased the step-like behaviour, like that we have
been discussing, is increasingly suppressed. For the single
component gas, we expect that an analysis similar to that given by
Luttinger~\cite{Luttinger} in the \dH effect could be used to show
that both the amplitude and period of the oscillations are not
affected by interactions to leading order. (This conclusion does
not hold for the non-oscillatory part of $\Omega$, that we have
called $\Omega_0$.) It would be of interest to study the Luttinger
analysis for a multi-component Fermi gas in more detail to make
contact with the results of Ref.~\cite{BB}.


\begin{thebibliography}{99}
\bibitem{BEC}
M. H. Anderson, J. R. Ensher, M. R. Matthews, C. E. Wieman, and E.
A. Cornell, Science {\bf 269}, 198 (1995); C. C. Bradley, C. A.
Sackett, J. J. Tollett, and R. G. Hulet, Phys. Rev. Lett. {\bf
75}, 1687 (1995); K. B. Davis, M.-O. Mewes, M. R. Andrews, N. J.
van Druten, D. S. Durfee, D. M. Kurn, and W. Ketterle, Phys. Rev.
Lett. {\bf 75}, 3969 (1995).

\bibitem{PethickSmith}
C.~J.~Pethick and H.~Smith, {\em Bose-Einstein Condensation in
Dilute Gases}, (Cambridge University Press, Cambridge, 2002).

\bibitem{fermi}
B. DeMarco and D. S. Jin, Science {\bf 285}, 1703 (1999); A. G.
Truscott, K. E. Strecker, W. I. McAlexander, G. B. Partridge, and
R. G. Hulet, Science {\bf 291}, 2570 (2001); F. Schreck, L.
Khaykovich, K. L. Corwin, G. Ferrari, T. Bourdel, J. Cubizolles,
and C. Salomon, Phys. Rev. Lett. {\bf 87}, 080403 (2001); S. R.
Granade, M. E. Gehm, K. M. O'Hara, and J. E. Thomas, Phys. Rev.
Lett. {\bf 88}, 120405 (2002); Z. Hadzibabic, C. A. Stan, K.
Dieckmann, S. Gupta, M.W. Zwierlein, A. Gorlitz, and W. Ketterle,
Phys. Rev. Lett. {\bf 88}, 160401 (2002); G. Roati, F. Riboli, G.
Modugno, and M. Inguscio, Phys. Rev. Lett. {\bf 89}, 150403
(2002).

\bibitem{BR}
D. A. Butts and D. S. Rokhsar, Phys. Rev. A {\bf 55}, 4346 (1997).

\bibitem{SW}
J. Schneider and H. Wallis, Phys. Rev. A {\bf 57}, 1253 (1998).

\bibitem{BB}
G. M. Bruun and K. Burnett, Phys. Rev. A {\bf 58}, 2427 (1998).

\bibitem{Brosens}
F. Brosens, J. T. Devreese, and L. F. Lemmens, Phys. Rev. E {\bf
57}, 3871 (1998).

\bibitem{VMT}
P. Vignolo, A. Minguzzi, and M. P. Tosi, Phys. Rev. Lett. {\bf
85}, 2850 (2000).

\bibitem{Gleisberg}
F. Gleisberg, W. Wonneberger, U. Schl\"{o}der, and C. Zimmermann,
Phys. Rev. A {\bf 62}, 063602 (2000).

\bibitem{BvZ}
M. Brack and B. P. van Zyl, Phys. Rev. Lett. {\bf 86}, 1574
(2001).

\bibitem{Wang}
X. Z. Wang, J. Phys. A {\bf 35}, 9601 (2002).

\bibitem{vZetal}
B. P. van Zyl, R. K. Bhaduri, A. Suzuki, and M. Brack,
Phys. Rev. A {\bf 67}, 023609 (2003).

\bibitem{BM}
M. Brack and M. V. N. Murthy, J. Phys. A {\bf 36}, 1111 (2003).

\bibitem{vZ}
B. P. van Zyl, Phys. Rev. A {\bf 68}, 033601 (2003).

\bibitem{DJTpartial}
D. J. Toms, J. Phys. A {\bf 37}, 3111 (2004).

\bibitem{Landau}
L. Landau, Z. Phys. {\bf 64}, 629 (1930).

\bibitem{Ashcroft}
N.~W. Ashcroft and N. D. Mermin, {\em Solid State Physics}, (Holt,
Rinehart, and Winston, Philadelphia, 1976).

\bibitem{Sond}
E.~H. Sondheimer and A.~H. Wilson, Proc. Roy. Soc. (London) {\bf
A210}, 173 (1951).

\bibitem{Huang}
K.~Huang, {\em Statistical Mechanics}, (J. Wiley, New York, 1987).

\bibitem{Pathria}
R.~K.~Pathria, {\em Statistical Mechanics}, (Pergamon, London,
1972).

\bibitem{WW}
E.~T.~Whittaker\ and G.~N.~Watson, {\em A Course of Modern
Analysis}, (Cambridge University Press, Cambridge, 1928).

\bibitem{Kirsten}
K. Kirsten and D. J. Toms, Phys. Lett. A {\bf 222}, 148 (1996); K.
Kirsten and D. J. Toms, Phys. Rev. E {\bf 59}, 158 (1999).

\bibitem{Luttinger}
J. M. Luttinger, Phys. Rev. {\bf 121}, 1251 (1961).

\end{thebibliography}
\end{document}